\documentclass[11pt]{article}

\usepackage[T1]{fontenc}
\usepackage{graphicx}
\usepackage{bxtexlogo}
\usepackage{enumerate}
\usepackage{mathtools}
\usepackage{verbatim}
\usepackage{setspace}
\usepackage{amsmath, amssymb}
\usepackage{amsthm}
\usepackage[top=30truemm,bottom=30truemm,left=27truemm,right=27truemm]{geometry}

\newtheorem{theorem}{Theorem}

\providecommand{\keywords}[1]
{
  \small	
  \textbf{Keywords:\ } #1
}

\begin{document}
\title{A Faster Deterministic Approximation Algorithm for TTP-2}
%
%
%
%
%


\author{Yuga Kanaya\thanks{Graduate School of Science and Engineering, Hosei University. 3-7-2, Kajino-cho, Koganei-shi, Tokyo 184-8584, Japan. \texttt{yuga.kanaya.4r@stu.hosei.ac.jp}} \and Kenjiro Takazawa\thanks{Faculty of Science and Engineering, Hosei University. 3-7-2, Kajino-cho, Koganei-shi, Tokyo 184-8584, Japan. \texttt{takazawa@hosei.ac.jp}}}
%

\date{}

\maketitle  
\begin{abstract}
The traveling tournament problem (TTP) is to minimize the total traveling distance of all teams in a double round-robin tournament.
In this paper, we focus on TTP-2, in which each team plays at most two consecutive home games and at most two consecutive away games.
For the case where the number of teams $n\equiv2$ (mod 4), Zhao and Xiao (2022) presented a $(1+5/n)$-approximation algorithm.
This is a randomized algorithm running in $O(n^3)$ time, and its derandomized version runs in $O(n^4)$ time.
In this paper, we present a faster deterministic algorithm running in $O(n^3)$ time, with approximation ratio $1+9/n$.
This ratio improves the previous approximation ratios of the deterministic algorithms with 
the same 
time complexity.

\end{abstract}
\section{Introduction}
The \textit{traveling tournament problem} (\textit{TTP}) is a major topic in sports scheduling.
The objective of this problem to minimize 
the total 
traveling distance of all teams in a double round-robin tournament of $n$ teams $t_1,\ldots,t_n$.
In a double round-robin tournament, each team will play two games 
against 
each of the other teams, one at its home venue and the other at its opponent's home venue.

In TTP, the tournament must satisfy the following four constraints.
\begin{itemize}
\item \textit{Fixed-game-time}: Each team will play one game in a day.
All the games should be scheduled in $2(n-1)$ consecutive days.
\item \textit{No-repeat}: Each team does not play 
against 
the same opponent in two consecutive games.
\item \textit{Start-and-end-points}: Each team stays at 
the 
home venue before the tournament, and returns to 
the 
home venue after the tournament.
\item \textit{Direct-travelng}: If team $t_{i}$ plays a game at team $t_{j}$'s home venue and the next game at team $t_{k}$'s home venue, team $t_{i}$ 
directly 
travels from $t_{j}$'s home venue to $t_{k}$'s home venue without returning to its home venue.
\end{itemize}
Let $k$ be a positive integer.
TTP with the following constraint is referred to as \textit{TTP-$k$}.
\begin{itemize}
\item \textit{Bounded-by-k}: Each team plays at most $k$ consecutive home games and at most $k$ consecutive away games.
\end{itemize}

For $i,j=1,\ldots,n$, let $d_{i,j}$ denote the distance from $t_{i}$'s home venue to $t_{j}$'s home venue.
We assume that the distances are metric, i.e., 
$d_{i,j}\ge 0$, 
$d_{i,i}=0$, 
$d_{i,j}=d_{j,i}$, and 
$d_{i,j}\leq d_{i,k}+d_{k,j}$ for each $i,j,k=1,\ldots,n$.


TTP-\textit{k} has been studied actively since a primary work of Easton \textit{et al.}\ \cite{Easton}.
For $k=1$, it is trivial that TTP-1 has no feasible schedule \cite{Werra}.
For $k=3$, Miyashiro \textit{et al.}\ \cite{Miyashiro} first designed a randomized approximation algorithm and then derandomized it.
The approximation ratio of these algorithms is $2+O(1/n)$.
It was improved to $5/3+O(1/n)$ by Yamaguchi \textit{et al.}\ \cite{Yamaguchi} and to $139/87+O(1/n)$ by Zhao \textit{et al.}\ \cite{ZhaoXiaoXu}.
For $k=4$, a $(17/10+O(1/n))$-approximation algorithm was designed \cite{ZhaoXiaoXu}.
For $k\geq5$, a $(5k-7)/(2k)+O(k/n)$-approximation algorithm was given \cite{Yamaguchi}.
For the case where $k\geq n-1$, that is, the bounded-by-$k$ constraint is ineffective, a 2.75-approximation algorithm was provided by Imahori \textit{et al.}\ \cite{ImahoriMatsuiMiyashiro}.
A $5$-approximation algorithm for all $k$ and a $4$-approximation algorithm for $k\geq n/2$ were designed \cite{ZhaoXiaoall}.

For $k=2$, all known algorithms for TTP-2 apply to either the case of $n\equiv0$ (mod 4) 
or that of $n\equiv2$ (mod 4), 
because the structural properties of the problem are different.
Two approximation algorithms for $n\equiv0$ (mod 4) and $n\equiv2$ (mod 4) were proposed by Thielen and Westphal \cite{ThielenWestphal}, who made a significant contribution to TTP-2.
Their approximation ratios are $1+16/n$ for $n\equiv0$ (mod 4) and $3/2+6/n$ for $n\equiv2$ (mod 4).
For $n\equiv0$ (mod 4), the approximation ratio was improved to $1+4/n$ by Xiao and Kou \cite{XiaoKou}, and an algorithm with approximation ratio $1+\frac{\lceil \log_2{n} \rceil+2}{2(n-2)}$, which is better for $n\leq32$, was also provided \cite{Chatterjee}.
Zhao and Xiao \cite{ZhaoXiao} further improved the approximation ratio to $(1+3/n)$.

For $n\equiv2$ (mod 4), 
Imahori \cite{Imahori}  presented an algorithm improving the approximation ratio from $3/2+6/n$ to $1+24/n$ for $n\geq30$.
This 
is a deterministic algorithm which 
uses a minimum weight perfect matching in numbering the teams and runs in $O(n^3)$ time. 
It is followed by a $(1+12/n)$-approximation algorithm by Zhao and Xiao \cite{ZhaoXiao2}, which also runs in $O(n^3)$ time.
More recently, Zhao and Xiao  \cite{ZhaoXiao3} 
further improved the approximation ratio to $1+5/n$. 
This improvement is based on
an elaborately constructed 
tournament in which each team plays as many two consecutive home games and away games as possible.
They designed a randomized algorithm running in $O(n^3)$ time, 
and its derandomized version running in $O(n^4)$ time.
The difference in the complexity 
comes from 
the fact that the
randomized algorithm randomly determines the numbering, 
while 
the derandomized one 
actually computes 
the optimal numbering.

In this paper, we present a new deterministic approximation algorithm for TTP-2 for $n\equiv2$ (mod 4).
The algorithm is developed by applying the numbering of the teams and 
estimation of the traveling distances 
by Imahori \cite{Imahori} to the tournament constructed by Zhao and Xiao \cite{ZhaoXiao3}.
We modify the numbering by Imahori \cite{Imahori} so that it can be applied to the tournament by Zhao and Xiao \cite{ZhaoXiao3}, 
and obtain new upper bounds (\ref{number1-2})--(\ref{number2}) on some of the traveling distances.
In addition, we estimate the traveling distance of each team in a way 
similar to that in \cite{Imahori}, 
which is 
different from that in \cite{ZhaoXiao3}.
Consequently, our algorithm has approximation ratio $1+9/n$ and time complexity $O(n^3)$.
This approximation ratio is better than those of 
the 
previous deterministic algorithms with the same complexity \cite{Imahori,ZhaoXiao2}.


\section{Preliminaries}
\subsection{Lower Bound}
In the analysis of the approximation ratio of our algorithm, we will use 
the following lower bound
on the total traveling distance \cite{CampbellChen}. 
Let $G=(V,E)$ denote the complete graph on $n$ vertices representing the $n$ teams, i.e., $V=\{t_1,\ldots,t_n\}$.
The weight of the edge between two vertices $t_{i}$ and $t_{j}$ is defined as the distance $d_{i,j}$.
Let $M\subseteq E$ denote a minimum weight perfect matching in $G$, 
and let $d(M)$ denote the sum of the weights of the edges in $M$.
Let $D(i)$ denote the total weight of the edges incident to $t_{i}$, i.e., $D(i)=\sum_{j\neq i}d_{i,j}$, and let $\Delta=\sum_{i=1}^n D(i)$.

For the away games of team $t_i$ against teams $t_j$ and $t_k$, the traveling distance of $t_i$ becomes smaller by playing these games in two consecutive days than returning to its home venue between them, because of the triangle inequality 
$d_{j,k}\leq d_{i,j}+d_{i,k}$.
Hence, $t_{i}$ should play as many two consecutive away games as possible.
Namely, $t_i$ plays two consecutive away games $(n/2-1)$ times and one away game between home games.
The total traveling distance of $t_{i}$ is minimized if each of the $(n/2-1)$ pairs of the two opponents in the consecutive away games form a pair in the minimum weight perfect matching $M$, and the opponent in the unique away game between home games is the team matched with $t_i$ in $M$.
The set of the edges in $G$ 
corresponding to this traveling is  
$\{(t_i,t_j)\mid j=1,2,\ldots,n,\ j\neq i\}\cup M$, where the edge in $M$ incident to $t_i$ is traversed twice.
Thus, $D(i)+d(M)$ is a lower bound on the total traveling distance of $t_{i}$.
Therefore, 
\begin{align}
\sum_{i=1}^{n}(D(i)+d(M))=\Delta+n\cdot d(M) \label{LB1}
\end{align}
is a lower bound on the total traveling distance of all teams.


\subsection{Construction of the Tournament by Zhao and Xiao} \label{Construction}
We describe the construction of the tournament by Zhao and Xiao \cite{ZhaoXiao3}.
Let $m=n/2$.
Recall that $M\subseteq E$ is a minimum weight perfect matching in $G$, and let $M=\{\{t_{2i-1},t_{2i}\}\mid i=1,\ldots,m\}$.
We define a \textit{super-team} $u_{i}$ as the pair of the endpoints of the edge $\{t_{2i-1},t_{2i}\}$ in $M$, i.e., $u_{i}=\{t_{2i-1}, t_{2i}\}\ (i=1,\ldots,m)$.

Each team plays $2n-2$ games, which are divided into $m-2$ \textit{super-games} 
and six games.
In each super-game, a team plays four games against teams in two or three super-teams, defined in the following way.

\begin{figure}
\centering
\includegraphics[width=0.55\linewidth]{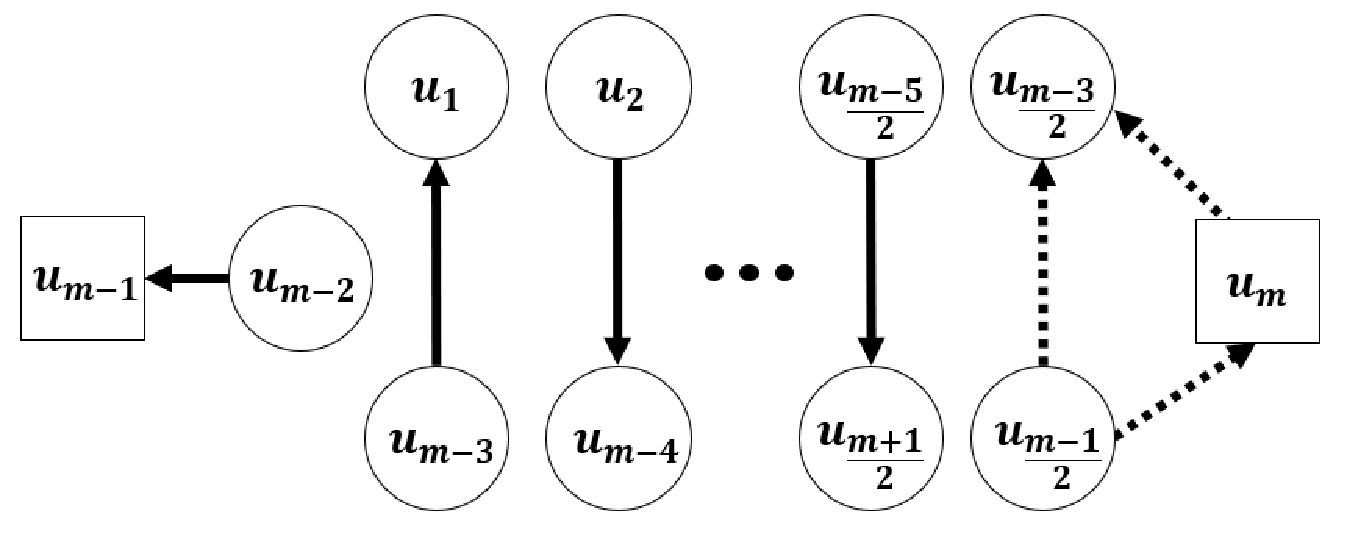}
\caption{Arrangement of super-teams in the first slot when $n\equiv2$ (mod 8)}
\label{fig2}
\end{figure}

In the first slot, the super-teams are arranged as shown in Figure \ref{fig2} if $n\equiv2$ (mod 8).
For the case $n\equiv6$ (mod 8), the three dotted edges are reversed (the rightmost straight edge ($(u_{(m-5)/2},u_{(m+1)/2})$ is also reversed).
The two and three super-teams connected by the directed edges form a super-game.
The super-game composed of three super-teams on the right is called a \textit{right super-game}, and the other super-games of two super-teams are called \textit{normal super-games}.
The direction of the edge(s) in a super-game determines the games in the super-game, which will be explained below.

After the first slot, all super-teams except for $u_{m-1}$ and $u_{m}$, i.e., the super-teams shown by circles in Figure \ref{fig2}, change the positions by moving one position in the clockwise direction, and all directed edges 
except for the leftmost edge incident to $u_{m-1}$
are reversed.
Consequently, in the second slot, the super-teams are arranged as shown in Figure \ref{fig3} in Appendix \ref{SECappA}.
The remaining slots are defined in the same manner.
One exception is that the direction of the leftmost edge is reversed in each slot.
The arrangement in the third slot is shown in Figure \ref{fig4} 
in Appendix \ref{SECappA}, 
and in the $(m-2)$-th slot in Figure \ref{fig5} in Appendix \ref{SECappA}.
From the second slot to the $(m-2)$-th slot, the super-game including $u_{m-1}$ is called a \textit{left super-game}
, represented by $L$ in Figure \ref{fig3} and Figure \ref{fig4}
in Appendix \ref{SECappA}.
Hence, in these slots, there are one left super-game, $(m-1)/2-2$ normal super-games and one right super-game.
The left super-game in the $(m-2)$-th slot 
is particularly 
called a \textit{special left super-game}
, represented by $S$ in Figure \ref{fig5} in Appendix \ref{SECappA}.

At the end of the $(m-2)$-th slot,
each team $t_i$ ($i=1,\ldots,n$) has played two games against $n-4$ teams and has played no game against 
the other three teams.
For $i=1,\ldots,m-2$,  
teams $t_{2i-1}$ and $t_{2i}$ in a super-team $u_i$ 
have played no 
game against each other, and against two of the four teams in the adjacent two super-teams.
Teams $t_{n-3}$, $t_{n-2}$, $t_{n-1}$ and $t_{n}$ 
have played no game 
against the three of these four teams except for itself.
Games against those three teams are played in 
the $(m-1)$-th
slot, which is in particular called \textit{the last slot}, 
consisting of six days.

Next, we show how to set up the games in the super-games.

\begin{figure}
\centering
\includegraphics[width=0.55\linewidth]{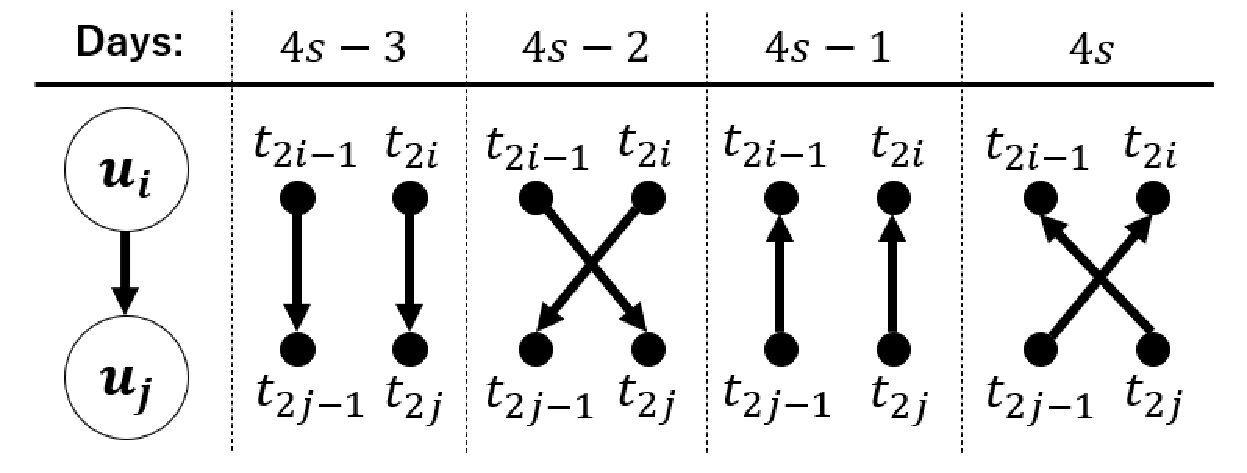}
\caption{The eight games in a normal super-game}
\label{fig6}
\end{figure}

\paragraph{Normal super-games:}
Each normal super-game is composed of eight games on four days.
Let a normal super-game of super-teams $u_i$ and $u_j$ is held in the $s$-th slot, where the direction of the edge is from $u_i$ to $u_j$ $(1\leq i, j\leq m-1$ and $1\leq s \leq m-2)$.
Recall that the super-team $u_{i}$ consists of two teams $t_{2i-1}$ and $t_{2i}$, and $u_{j}$ of $t_{2j-1}$ and $t_{2j}$.
The eight games in this super-game will be played in days from $4s-3$ to $4s$, as shown in Figure \ref{fig6}.
In Figure \ref{fig6}, each directed edge represents a game held in the home venue of the team of the tail vertex.

\begin{figure}
\centering
\includegraphics[width=0.55\linewidth]{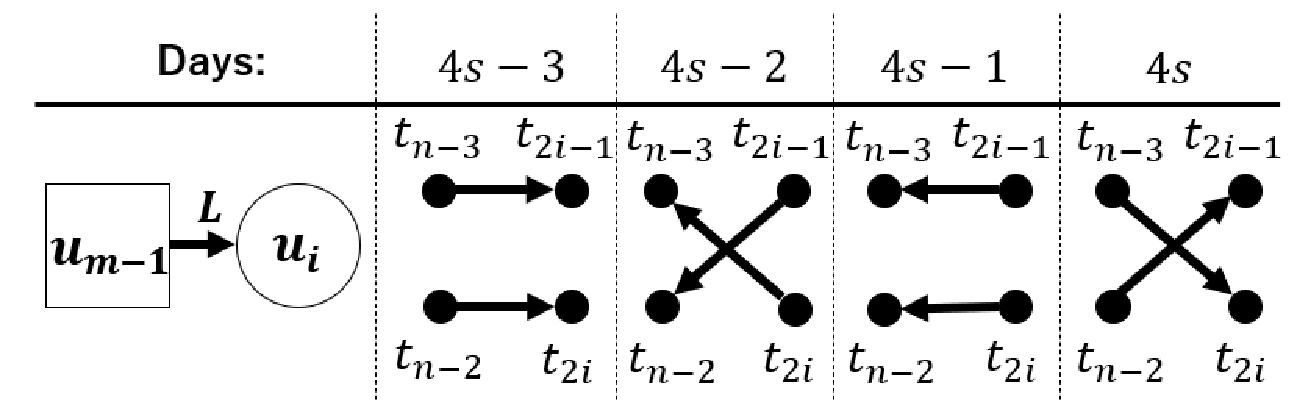}
\caption{The eight games in a left super-games}
\label{fig7}
\end{figure}

\paragraph{Left super-games:}
Each left super-game is composed of eight games on four days.
Let a left super-game of super-teams $u_{i}$ and $u_{m-1}$ is held in the $s$-th slot 
($2\leq i \leq m-3$, $2\leq s\leq m-3$, $i+s=m-1$), 
and 
suppose that the direction of the edge is from $u_{i}$ to $u_{m-1}$.
Recall that the super-team $u_{m-1}$ consists of two normal teams $t_{n-3}$ and $t_{n-2}$.
The eight games in this super-game will be played in days from $4s-3$ to $4s$, as shown in Figure \ref{fig7}.
If directed edge between two super-teams is reversed, all directed edge between two teams are reversed.

\begin{figure}
\centering
\includegraphics[width=0.55\linewidth]{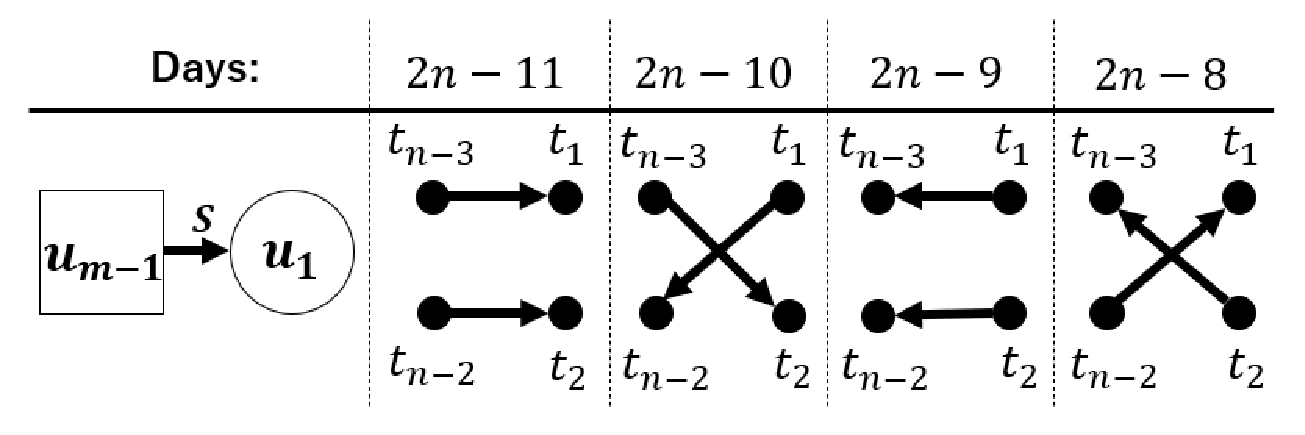}
\caption{The eight games in the special left super-game in the $(m-2)$-th slot}
\label{fig8}
\end{figure}

The special left super-game in the $(m-2)$-th slot consists of super-teams $u_1$ and $u_{m-1}$.
The eight games in this super-game will be played in days from $2n-11$ to $2n-8$, as shown in Figure \ref{fig8}.
Special left super-game differs from other left super-games in that the edges incident to team $t_{n-3}$ on days $2n-10$ and $2n-8$
are reversed. 


\paragraph{Right super-games:}
A right super-game consists of twelve games on four days, played by three super-teams (six teams).
For concise notation, let $u_{0}=u_{m-2}$.
Let the super-game be held in the
$s$-th slot, and denote the three super-teams by $u_{i-1}$, $u_{i}$ and $u_{m}$ 
($1\leq i \leq m-2,\ 1\leq s \leq m-2,\ i+s\equiv(m+1)/2$ (mod $m-2$)). 
Suppose that the direction of the edges are from $u_{i-1}$ to $u_{i}$, from $u_{i-1}$ to $u_{m}$ and from $u_{m}$ to $u_{i}$.

\begin{figure}
\centering
\includegraphics[width=0.5\linewidth]{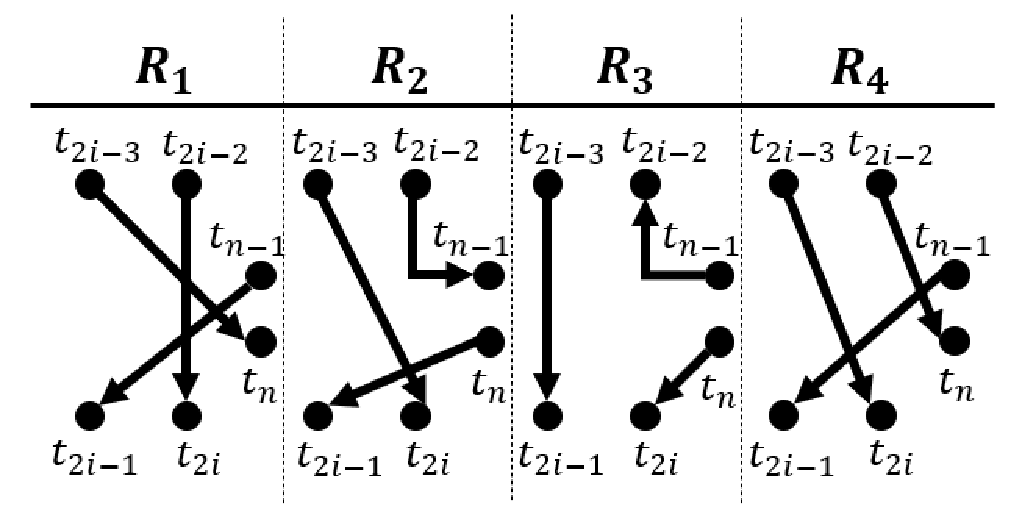}
\caption{Four types of right super-games}
\label{fig9}
\end{figure}

The three games in each day are of eight types, $R_1,\ldots,R_4,\overline{R_1},\ldots,\overline{R_4}$, shown in Figure \ref{fig9}.
For $\ell=1,\ldots,4$, the type $\overline{R_\ell}$ is obtained from $R_\ell$ by reversing the edges.
For $1\leq s \leq (m-3)/2$, if there is a directed edge from $u_{i-1}$ to $u_{i}$, the four days are arranged as $R_4, R_3, \overline{R_4}, \overline{R_3}$, and otherwise $\overline{R_2}, \overline{R_1}, R_2, R_1$.
For $s=(m-1)/2$, there is a directed edge from $u_{m-2}$ to $u_{1}$, and the four days are arranged as $R_1, R_3, \overline{R_1}, \overline{R_3}$.
For $(m+1)/2\leq s \leq m-2$, if there is a directed edge from $u_{i-1}$ to $u_{i}$, the four days are arranged as $R_1, R_2, \overline{R_1}, \overline{R_2}$, and otherwise $\overline{R_3}, \overline{R_4}, R_3, R_4$.

\begin{figure}
\centering
\includegraphics[width=0.5\linewidth]{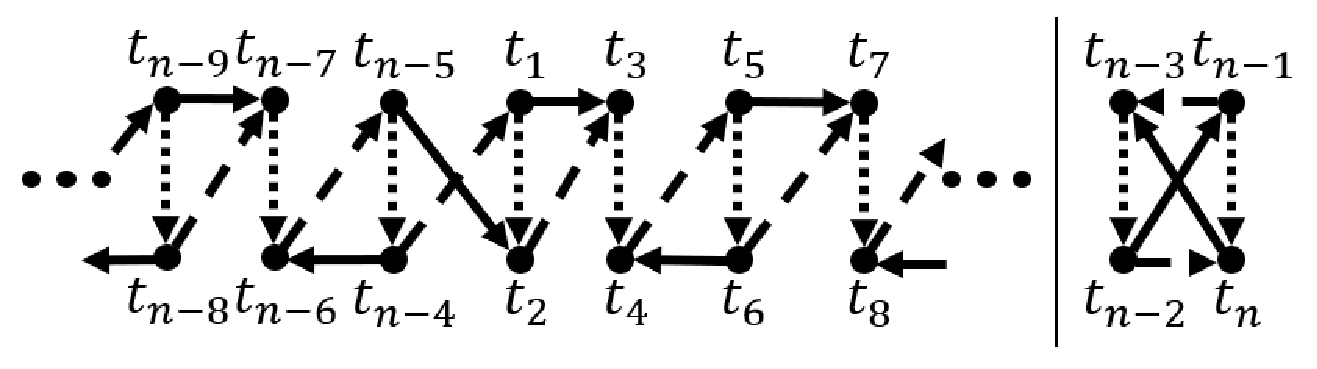}
\caption{The first three  
days 
in the last slot.}
\label{fig10}
\end{figure}

\begin{figure}
\centering
\includegraphics[width=0.5\linewidth]{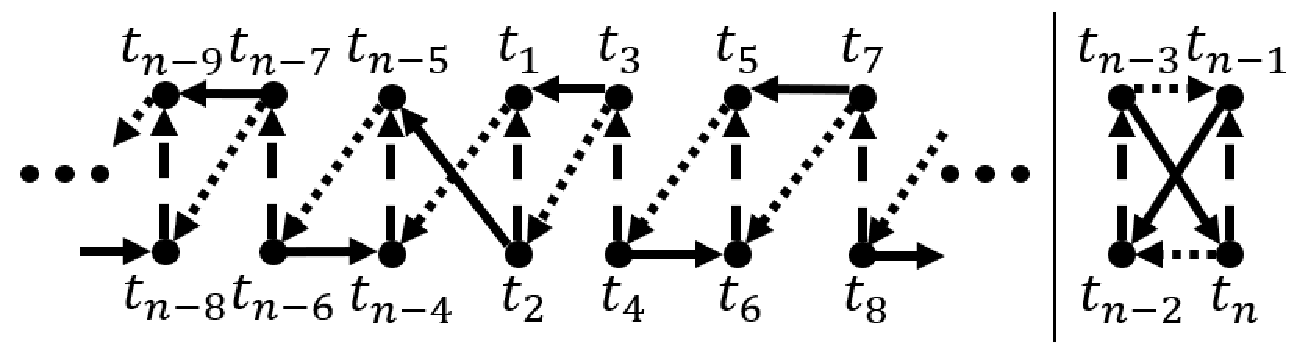}
\caption{The last three 
days
in the last slot.}
\label{fig11}
\end{figure}

\paragraph{The last slot:}
The first three 
days 
are shown in Figure \ref{fig10}, and the last three 
days 
in Figure \ref{fig11}.
Figures 
\ref{fig10} and \ref{fig11} have directed edges of three types.
A straight directed edge represents a game on the first day in Figure \ref{fig10} and on the fourth day in Figure \ref{fig11}.
A dotted directed edge represents a game on the second day in Figure \ref{fig10} and on the fifth day in Figure \ref{fig11}.
A broken directed edge represents a game on the third day in Figure \ref{fig10} and on the sixth day in Figure \ref{fig11}.

\section{Our Algorithm}
In this section we present our algorithm.
The algorithm is designed by combining the construction of the tournament by Zhao and Xiao \cite{ZhaoXiao3} described in Section \ref{Construction} and the numbering of the teams based on the idea of Imahori \cite{Imahori}.
\subsection{Numbering of the Teams}
Recall that $M$ is a minimum weight perfect matching in $G$ and 
the two teams $t_{2i-1}$ and $t_{2i}$ must be connected by an edge in $M$ so that they form a super-team $u_i$ ($i=1,\ldots, m$).
We determine four teams $t_{n-3},t_{n-2},t_{n-1}$, and $t_{n}$, 
forming two super-teams $u_{m-1}$ and $u_m$, 
so that $D(n-1)+D(n)$ is the smallest and $D(n-3)+D(n-2)$ is the second smallest of $D(2i-1)+D(2i)$ $(i=1,2,\ldots,m)$. 
The other teams $t_1,\ldots, t_{n-4}$ are numbered arbitrarily as long as 
$\{t_{2i-1} , t_{2i}\}$ is an edge in $M$ for each $i=1,\ldots, m-2$. 



From
this numbering, 
we obtain the following three inequalities: 
\begin{align}
&{}D(n-1)+D(n)\leq \frac{2}{n}\Delta \label{number1-2}, \\
&{}D(n-3)+D(n-2)+D(n-1)+D(n)\leq \frac{4}{n}\Delta, \label{number1}\\
&{}d_{2,4}+d_{4,6}+\cdots+d_{n-6,n-4}+d_{n-4,2}\leq \frac{1}{n}(\Delta+n\cdot d(M)). \label{number2}
\end{align}
Inequalities \eqref{number1-2} and \eqref{number1}
are derived from the fact that 
$D(n-1)+D(n)$ is the smallest and $D(n-3)+D(n-2)$ is the second smallest $D(2i-1)+D(2i)$ $(i=1,2,\ldots,m)$.
We derive \eqref{number2} 
by using the triangle inequality and (\ref{number1-2}) as 
\begin{align}
&d_{2,4}+d_{4,6}+\cdots+d_{n-6,n-4}+d_{n-4,2} \notag
\\ &\leq (d_{2,3}+d_{3,4})+(d_{4,5}+d_{5,6})+\cdots+(d_{n-6,n-5}+d_{n-5,n-4})+(d_{n-4,1}+d_{1,2})\notag
\\ &= (d_{2,3}+d_{4,5}+\cdots+d_{n-6,n-5}+d_{n-4,1})+(d_{1,2}+d_{3,4}+\cdots+d_{n-5,n-4})\notag
\\ &\leq \sum_{j=1}^{m-3}d_{2j,2j+1}+d_{n-4,1}+d(M)\notag
\\ &\leq \frac{1}{2}\left(\sum_{j=1}^{m-3}(d_{2j,n-1}+d_{2j+1,n-1}+d_{2j,n}+d_{2j+1,n})+
d_{n-4,n-1}+d_{1,n-1}+d_{n-4,n}+d_{1,n}\right)+d(M)\notag
\\ &\leq \frac{1}{2}\sum_{i=1}^{n-4}(d_{i,n-1}+d_{i,n})+d(M)\notag
\\ &\leq \frac{1}{2}(D(n-1)+D(n))+d(M)\notag
= \frac{1}{n}(\Delta+n\cdot d(M)).\notag
\end{align}

\subsection{Analyzing the Approximation Ratio} \label{approximation ratio}
Here we describe the analysis of the approximation ratio in the case of $n\equiv2$ (mod 8). 
The case of $n\equiv6$ (mod 8) can be analyzed in the same way.

Recall the lower bound (\ref{LB1}), 
and define an \emph{extra traveling} as a traveling of an edge that is not counted in defining (\ref{LB1}).
We estimate the distances of the extra travelings 
according to the following categorization of the super-teams into six classes.

\paragraph{Category $\mathrm{(I)}$: Super-team $u_{1}$.}
Teams $t_1$ and $t_2$ in super-team $u_{1}$ will play $m-5$ normal super-games, one special left super-game, two right super-games, and six games in the last slot.
The travels of teams $t_{1}$ and $t_{2}$ are shown in Figures \ref{fig12}--\ref{fig15} in Appendix \ref{SECappA}.
In the $m-5$ normal super-games and the special left super-game, teams $t_1$ and $t_2$ have no extra traveling.
The distance of the extra traveling of $t_1$ in the two right super-games is $d_{4,n}+d_{n-5,n-1}$, and that of $t_2$ is $d_{4,n-1}+d_{n-4,n}$.
The distance of the extra traveling of $t_1$ in the six games in the last slot is $d_{2,3}+d_{1,n-4}$, and that of $t_2$ is $d_{3,n-5}+d_{1,2}$.
These extra traveling distances are summarized in Table \ref{table1}.
\begin{table}[t]
\centering
 \caption{Extra traveling distances of $t_1$ and $t_2$}
  \begin{tabular}{|l|c|c|}
  \hline
   & $t_{1}$ & $t_{2}$ \\
  \hline
  Right super-games & \multicolumn{1}{l|}{\textbf{(I-a)} $d_{4,n}+d_{n-5,n-1}$} & \multicolumn{1}{l|}{\textbf{(I-b)} $d_{4,n-1}+d_{n-4,n}$} \\
  Six games in the last slot & \multicolumn{1}{l|}{\textbf{(I-c)} $d_{2,3}+d_{1,n-4}$} & \multicolumn{1}{l|}{\textbf{(I-d)} $d_{3,n-5}+d_{1,2}$} \\
  \hline
  \end{tabular}
  \label{table1} 
\end{table}

\paragraph{Category $\mathrm{(II)}$: Super-teams $u_{i}$ $(i=2,4,\ldots,m-3)$.}
Teams $t_{2i-1}$ and $t_{2i}$ in super-team $u_{i}$ $(i=2,4,\ldots,m-3)$ will play $m-5$ normal super-games, one left super-game, two right super-games, and six games in the last slot.
The travels of teams $t_{2i-1}$ and $t_{2i}$ are shown in Figure \ref{fig16}--\ref{fig19} in Appendix \ref{SECappA}.
The distances of the extra traveling are summarized in Table \ref{table2}.
\begin{table}[t]
 \centering
 \caption{Extra traveling distances of $t_{2i-1}$ and $t_{2i}$ ($i=2,4,\ldots,m-3$)}
  \begin{tabular}{|l|c|c|}
  \hline
   & $t_{2i-1}$ & $t_{2i}$ \\
  \hline
  Left & \multicolumn{1}{l|}{\textbf{(II-a)} $d_{2i-1,n-3}+d_{2i-1,n-2}$} & \multicolumn{1}{l|}{\textbf{(II-b) $d_{2i,n-3}+d_{2i,n-2}$}} \\
  Right & \multicolumn{1}{l|}{\textbf{(II-c)} $d_{2i+1,2i+2}$} & \multicolumn{1}{l|}{\textbf{(II-d)} $d_{2i-3,2i-2}$} \\
  Six games & \multicolumn{1}{l|}{\textbf{(II-e)} $d_{2i-1,2i}+d_{2i-3,2i-2}$} & \multicolumn{1}{l|}{\textbf{(II-f)} $d_{2i+1,2i+2}+d_{2i-1,2i}$} \\
  \hline
    \end{tabular}
  \label{table2} 
\end{table}

\paragraph{Category $\mathrm{(III)}$: Super-teams $u_{i}$ $(i=3,5,\ldots,m-4)$.}
Teams $t_{2i-1}$ and $t_{2i}$ in super-team $u_{i}$ $(i=3,5,\ldots,m-4)$ will play $m-5$ normal super-games, one left super-game, two right super-games, and six games in the last slot.
The travels of teams $t_{2i-1}$ and $t_{2i}$ are shown in Figure \ref{fig20}--\ref{fig23} in Appendix \ref{SECappA}.
The distances of the extra traveling are summarized in Table \ref{table3}.
\begin{table}[t]
 \centering
 \caption{Extra traveling distances of $t_{2i-1}$ and $t_{2i}$ ($i=3,5,\ldots,m-4$)}
  \begin{tabular}{|l|c|c|}
  \hline
   & $t_{2i-1}$ & $t_{2i}$ \\
  \hline
  Right & \multicolumn{1}{l|}{\textbf{(III-a)} $d_{2i+2,n}+d_{2i-3,n-1}$} & \multicolumn{1}{l|}{\textbf{(III-b)} $d_{2i+2,n-1}+d_{2i-3,n}$} \\
  Six games & \multicolumn{1}{l|}{\textbf{(III-c)} $d_{2i,2i+1}+d_{2i-2,2i-1}$} & \multicolumn{1}{l|}{\textbf{(III-d)} $d_{2i-2,2i}+d_{2i,2i+1}+d_{2i-1,2i}$} \\ 
  \hline
  \end{tabular}
 \label{table3} 
\end{table}

\paragraph{Category $\mathrm{(IV)}$: Super-team $u_{m-2}$.}
Teams $t_{n-5}$ and $t_{n-4}$ in super-team $u_{m-2}$ will play $m-4$ normal super-games, two right super-games, and six games in the last slot.
The travels of teams $t_{n-5}$ and $t_{n-4}$ are shown in Figure \ref{fig24}--\ref{fig26} in Appendix \ref{SECappA}.
The distances of the extra traveling are summarized in Table \ref{table4}.
\begin{table}[t]
 \centering
 \caption{Extra traveling distances of $t_{n-5}$ and $t_{n-4}$}
  \begin{tabular}{|l|c|c|}
  \hline
   & $t_{n-5}$ & $t_{n-4}$ \\
  \hline
  Right & \multicolumn{1}{l|}{\textbf{(IV-a)} $d_{1,n}+d_{n-7,n-1}$} & \multicolumn{1}{l|}{\textbf{(IV-b)} $d_{2,n-1}+d_{n-7,n}$} \\
  Six games & \multicolumn{1}{l|}{\textbf{(IV-c)} $d_{n-4,2}+d_{n-6,n-5}$} & \multicolumn{1}{l|}{\textbf{(IV-d)} $d_{n-6,n-4}+d_{1,n-4}+d_{n-5,n-4}$} \\
  \hline
  \end{tabular}
 \label{table4} 
\end{table}  

\paragraph{Category $\mathrm{(V)}$: Super-team $u_{m-1}$.}
Teams $t_{n-3}$ and $t_{n-2}$ in super-team $u_{m-1}$ will play one normal super-game, $m-3$ left super-games, and six games in the last slot.
The travels of teams $t_{n-3}$ and $t_{n-2}$ are shown in Figure \ref{fig27}--\ref{fig30} in Appendix \ref{SECappA}.
The distances of the extra traveling are summarized in Table \ref{table5}.
\begin{table}[t]
 \centering
 \caption{Extra traveling distances of $t_{n-3}$ and $t_{n-2}$}
  \begin{tabular}{|l|c|c|}
  \hline
   & $t_{n-3}$ & $t_{n-2}$ \\
   \hline
  Left & \multicolumn{1}{l|}{\textbf{(V-a)}} & \multicolumn{1}{l|}{\textbf{(V-b)}} \\
   & \multicolumn{1}{l|}{\ $\displaystyle{\sum_{i=1}^{(n-10)/4}(d_{4i+1,n-3}+d_{4i+2,n-3})}$} & \multicolumn{1}{l|}{\ $\displaystyle{\sum_{i=1}^{(n-10)/4}(d_{4i+1,n-2}+d_{4i+2,n-2})}$} \\
   &  & \multicolumn{1}{l|}{\ $+d_{1,n-2}+d_{2,n-2}$} \\
 \hline
  Six games & \multicolumn{1}{l|}{\textbf{(V-c)}} & \multicolumn{1}{l|}{\textbf{(V-d)}} \\
   & \multicolumn{1}{l|}{\ $d_{n-2,n-3}+d_{n,n-1}$} & \multicolumn{1}{l|}{\ $d_{n-1,n}+d_{n-3,n-2}$} \\
   \hline
  \end{tabular}
 \label{table5} 
\end{table}   

\paragraph{Category $\mathrm{(VI)}$: Super-team $u_{m}$.}
Teams $t_{n-1}$ and $t_{n}$ in super-team $u_{m}$ will play $m-2$ right super-games, and six games in the last slot.
The travels of teams $t_{n-1}$ and $t_{n}$ are shown in Figure \ref{fig41}--\ref{fig36} in Appendix \ref{SECappA}.
The distances of the extra traveling are summarized in Table \ref{table6}.
\begin{table}[t]
 \centering
 \caption{Extra traveling distances of $t_{n-1}$ and $t_{n}$}
  \begin{tabular}{|l|c|c|}
  \hline
   & $t_{n-1}$ & $t_{n}$ \\
  \hline
  Right & \multicolumn{1}{l|}{\textbf{(VI-a)}} & \multicolumn{1}{l|}{\textbf{(VI-b)}} \\
   & \multicolumn{1}{l|}{\ $\displaystyle{\sum_{i=1}^{(m-5)/4}(d_{4i-3,4i-1}+d_{4i-2,4i})}$} & \multicolumn{1}{l|}{\ $\displaystyle{\sum_{i=1}^{(m-3)/2}(d_{4i-3,4i-1}+d_{4i,4i+2})}$} \\
   & \multicolumn{1}{l|}{\ $\displaystyle{+\sum_{(m+3)/4}^{(m-3)/2}(d_{4i-1,4i+1}+d_{4i,4i+2})}$} & \multicolumn{1}{l|}{\ $+d_{2,n-5}$} \\
   & \multicolumn{1}{l|}{\ $+d_{m-4,m-2}+d_{m-1,m+1}$} & \\
   & \multicolumn{1}{l|}{\ $+d_{m-3,n-1}+d_{m,n}$} & \\
   \hline
  Six games & \multicolumn{1}{l|}{\textbf{(VI-c)}} & \multicolumn{1}{l|}{\textbf{(VI-d)}} \\
   & \multicolumn{1}{l|}{\ $d_{n-3,n-2}+d_{n-1,n}$} & \multicolumn{1}{l|}{\ $d_{n-2,n-1}+d_{n-3,n}$} \\
   \hline
  \end{tabular}
 \label{table6} 
\end{table}     


\medskip

We have enumerated the distances of the extra travelings of all teams $t_{i}$.
We provide an upper bound of some of these distances by using triangle inequality, as shown Table \ref{table7}.
\begin{table}[h]
 \centering
 \caption{Upper bounds of some of extra traveling distances}
  \begin{tabular}{|c|c|c|}
  \hline
   & Distance & Upper bound\\
  \hline
  (I-c) & \multicolumn{1}{l|}{$d_{2,3}+d_{1,n-4}$} & \multicolumn{1}{l|}{\ $(d_{2,4}+d_{3,4})+(d_{n-4,2}+d_{1,2})$} \\\hline
  (I-d) & \multicolumn{1}{l|}{$d_{3,n-5}$} & \multicolumn{1}{l|}{\ $d_{3,4}+d_{4,6}+d_{6,n}+d_{n-5,n}$} \\\hline
  (III-c) & \multicolumn{1}{l|}{$d_{2i,2i+1}+d_{2i-2,2i-1}$} & \multicolumn{1}{l|}{\ $(d_{2i,2i+2}+d_{2i+1,2i+2})$} \\
   &  & \multicolumn{1}{l|}{\ $+(d_{2i-2,2i}+d_{2i-1,2i})$} \\\hline
  (III-d) & \multicolumn{1}{l|}{$d_{2i-2,2i}+d_{2i,2i+1}$} & \multicolumn{1}{l|}{\ $d_{2i-2,2i}+(d_{2i,2i+2}+d_{2i+1,2i+2})$} \\\hline
  (IV-c) & \multicolumn{1}{l|}{$d_{n-4,2}+d_{n-6,n-5}$} & \multicolumn{1}{l|}{\ $(d_{n-4,n-3}+d_{2,n-3})$} \\
   &  & \multicolumn{1}{l|}{\ $+(d_{n-6,n-4}+d_{n-5,n-4})$} \\\hline
  (IV-d) & \multicolumn{1}{l|}{$d_{n-6,n-4}+d_{1,n-4}$} & \multicolumn{1}{l|}{\ $d_{n-6,n-4}+(d_{1,n-1}+d_{n-4,n-1})$} \\\hline
  (VI-a) & \multicolumn{1}{l|}{$\displaystyle{\sum_{i=1}^{(m-5)/4}(d_{4i-3,4i-1}+d_{4i-2,4i})}$} & \multicolumn{1}{l|}{\ $\displaystyle{\sum_{i=1}^{(m-5)/4}\{(d_{4i-3,4i-2}+d_{4i-2,4i}}$} \\
   &  & \multicolumn{1}{l|}{\ $+d_{4i-1,4i})+d_{4i-2,4i}$\}} \\
   & \multicolumn{1}{l|}{$\displaystyle{+\sum_{i=(m+3)/4}^{(m-3)/2}(d_{4i-1,4i+1}+d_{4i,4i+2})}$} & \multicolumn{1}{l|}{\ $\displaystyle{+\sum_{i=(m+1)/4}^{(m-3)/2}\{(d_{4i-1,4i}+d_{4i,4i+2}}$} \\
    &  & \multicolumn{1}{l|}{\ $+d_{4i+1,4i+2})+d_{4i,4i+2}\}$} \\
   & \multicolumn{1}{l|}{$+d_{m-4,m-2}+d_{m-1,m+1}$} & \multicolumn{1}{l|}{\ $(d_{m-4,m-3}+d_{m-3,m-1}$} \\
    &  & \multicolumn{1}{l|}{\ $+d_{m-2,m-1})+d_{m-1,m+1}$} \\\hline
  (VI-b) & \multicolumn{1}{l|}{$\displaystyle{\sum_{i=1}^{(m-3)/2}(d_{4i-3,4i-1}+d_{4i,4i+2})}$} & \multicolumn{1}{l|}{\ $\displaystyle{\sum_{i=1}^{(m-3)/2}\{(d_{4i-3,4i-2}+d_{4i-2,4i}}$} \\
   &  & \multicolumn{1}{l|}{\ $+d_{4i-1,4i})+d_{4i,4i+2}\}$} \\
   & \multicolumn{1}{l|}{$+d_{2,n-5}$} & \multicolumn{1}{l|}{\ $+d_{n-5,n-4}+d_{n-4,2}$} \\\hline
  (VI-c) & \multicolumn{1}{l|}{$d_{n-3,n-2}+d_{n-1,n}$} & \multicolumn{1}{l|}{$(d_{n-1,n-3}+d_{n-1,n-2})$} \\
   &   & \multicolumn{1}{l|}{\ $+(d_{5,n-1}+d_{5,n})$} \\
   \hline
 \end{tabular}
 \label{table7} 
\end{table}
It follows that all terms of these upper bounds 
and 
the distances of the extra travelings covered by the upper bounds 
are of the following three types:
\begin{enumerate}[(A)]
\item $d_{i,n-3},d_{i,n-2},d_{i,n-1},d_{i,n}$ $(i=1,\ldots,n)$;
\item $d_{2,4},d_{4,6},\ldots,d_{n-6,n-4},d_{n-4,2}$; and
\item $d_{1,2},d_{3,4}\ldots,d_{n-1,n}$.
\end{enumerate}

For each of these three types, we provide
an upper bound of the sum of the distances 
in the following manner. 
First, we focus on Type (A).
The distances $d_{i,n-3}, d_{i,n-2}, d_{i,n-1}$ and $d_{i,n}$ $(i=1,\ldots,n)$ appear in (I-a), (I-b), (I-d), (II-a), (II-b), (III-a), (III-b), (IV-a), (IV-b), (IV-c), (IV-d), (V-a), (V-b), (V-c), (V-d), (VI-a), (VI-c), and (VI-d).
Observe that each of $d_{i,n-3}, d_{i,n-2}, d_{i,n-1}$ and $d_{i,n}$ $(i=1,\ldots,n)$ appears at most once, and hence the sum of the distances of Type (A) is at most
\begin{align}
D(n-3)+D(n-2)+D(n-1)+D(n). \label{total1}
\end{align}

Second, we focus on the distances of Type (B).
These appear in (I-c), (I-d), (III-c), (III-d), (IV-c), (IV-d), (VI-a), and (VI-b).
Observe that each of $d_{2,4},d_{4,6},\ldots,d_{n-6,n-4},d_{n-4,2}$ appears at most five times, and hence the sum of the distances of Type (B) is at most
\begin{align}
5(d_{2,4}+d_{4,6}+\cdots+d_{n-6,n-4}+d_{n-4,2}). \label{total2}
\end{align}

Last, we focus on the distances of Type (C).
These appear in (I-c), (I-d), (II-c), (II-d), (II-e), (II-f), (III-c), (III-d), (IV-c), (IV-d), (VI-a), and (VI-b).
Observe that each of $d_{1,2},d_{3,4},\ldots,d_{n-1,n}$ appears at most six times, and hence
the sum of the distances of Type (C) is at most
\begin{align}
6(d_{1,2}+d_{3,4}+\cdots+d_{n-1,n})=6\cdot d(M). \label{dm1}
\end{align}

We have provided upper bounds on the extra traveling distances.
Finally, we
investigate the distances counted in the lower bound (\ref{LB1}), because some of 
the 
edges in the perfect matching $M$ are not traveled 
by the two teams $t_{n-1}$ and $t_n$
in our tournament.
In the first $m-2$ slots, each of $t_{n-1}$ and $t_{n}$ plays right super-games and the two opponents in one right super-game are in different super-teams.
Hence, each of $d_{1,2},d_{3,4},\ldots,d_{n-1,n}$ does not appear in the traveling of $t_{n-1}$ and $t_{n}$.
Accordingly, the distances of the edges counted in the lower bound (\ref{LB1}) and used in our tournament is at most
\begin{align}
\Delta+n\cdot d(M)-2\cdot d(M)=\Delta+(n-2)d(M). \label{dm2}
\end{align}

Therefore, by summing (\ref{total1}), (\ref{total2}), (\ref{dm1}), and (\ref{dm2}), we obtain that the total traveling distance is at most
\begin{align*}
&\Delta+D(n-3)+D(n-2)+D(n-1)+D(n)\\
&+5(d_{2,4}+d_{4,6}+\cdots+d_{n-6,n-4}+d_{n-4,2})+
(n+4)\cdot d(M).
\end{align*}
It follows from (\ref{number1}) and (\ref{number2}) that it is at most 
\begin{align*}
&\Delta+\frac{4}{n}\Delta+5\cdot \frac{1}{n}(\Delta+n\cdot d(M))+
(n+4)\cdot d(M) \\
&=\left(1+\frac{9}{n}\right)(\Delta+n\cdot d(M)).
\end{align*}
Recall
that $\Delta+n\cdot d(M)$ is lower bound (\ref{LB1}).
We thus conclude that the approximation ratio of our algorithm is $1+9/n$.

\begin{theorem}
The approximation ratio of our algorithm is $1+9/n$.
\end{theorem}

\subsection{Analyzing the Time Complexity}
Recall that $G$ is a complete graph with $n$ vertices.
Let $L$ denote the maximum weight of an edge in $G$, and $\mu(n,L)$
the time complexity for finding a minimum weight perfect matching in $G$. 
Currently, 
$$\mu(n,L)=O(\min\{n^3,n^2(\log(nL)\sqrt{n\cdot \alpha(n^2,n)\cdot \log(n)})\})$$
where $\alpha(\cdot,\cdot)$ is the inverse Ackermann function \cite{PM1-1,PM2,PM1-2}. 

We first compute a minimum weight perfect matching $M$
in $G$.
We second compute the sum of $D(\cdot)$ of the two teams in each edge of $M$ to determine $u_{m}$ and $u_{m-1}$, which requires $O(n)$ time.
Thus, the computation time of $M$ dominates the other, and hence the computation time of our algorithm is 
$\mu(n,L)$,
which is $O(n^3)$.

\begin{theorem}
The time complexity of our algorithm is $O(n^3)$.
\end{theorem}

\section{Conclusion}
We have presented a deterministic $(1+9/n)$-approximation algorithm for 
TTP-2, where 
$n$ is the number of the teams satisfying $n\equiv2$ (mod 4).
This algorithm runs in $O(n^3)$ time.
Compared with the deterministic $(1+5/n)$-approximation algorithm by Zhao and Xiao \cite{ZhaoXiao3}, which runs in $O(n^4)$ time, our algorithm has better complexity.
Compared with the algorithm having the same complexity \cite{Imahori,ZhaoXiao2}, our algorithm has  
a better approximation ratio.

One direction of future work is to improve the approximation ratio.
There is a gap between the total traveling distance in our algorithm and its upper bound obtained in Section 3.2.
Possibly a 
further
analysis of the upper bound 
results in a better approximation guarantee of our algorithm.
Another direction is to investigate the time complexity of TTP-2.
For $k\geq3$, TTP-$k$ is shown to be NP-hard by Thielen and Westphal \cite{ThielenWestphal2}.
However, the complexity of TTP-2 is still not revealed, while a number of approximation algorithms are designed.



\bibliographystyle{abbrv} 
\bibliography{main.bib}

\begin{thebibliography}{10}

\bibitem{CampbellChen}
R.~T. Campbell and D.~Chen.
\newblock A minimum distance basketball scheduling problem.
\newblock {\em Management Science in Sports}, 4:15--26, 1976.

\bibitem{Chatterjee}
D.~Chatterjee and B.~K. Roy.
\newblock An improved scheduling algorithm for traveling tournament problem with maximum trip length two.
\newblock {\em arXiv:2109.09065}, 2021.

\bibitem{Werra}
D.~de~Werra.
\newblock Some models of graphs for scheduling sports competitions.
\newblock {\em Discrete Appl. Math.}, 21(1):47--65, 1988.

\bibitem{Easton}
K.~Easton, G.~Nemhauser, and M.~Trick.
\newblock The traveling tournament problem description and benchmarks.
\newblock In {\em 7th CP}, pages 580--584. Springer, 2001.

\bibitem{PM1-1}
H.~N. Gabow.
\newblock {\em Implementation of Algorithms for Maximum Matching on Nonbipartite Graphs}.
\newblock PhD thesis, Department of Computer Science, Stanford University, Stanford, California, 1973.

\bibitem{PM2}
H.~N. Gabow and R.~E. Tarjan.
\newblock Faster scaling algorithms for general graph matching problems.
\newblock {\em J. ACM}, 38(4):815--853, 1991.

\bibitem{Imahori}
S.~Imahori.
\newblock A 1+ $\mathrm{O}(1/n)$ approximation algorithm for $\mathrm{TTP}$ (2).
\newblock {\em arXiv:2108.08444}, 2021.

\bibitem{ImahoriMatsuiMiyashiro}
S.~Imahori, T.~Matsui, and R.~Miyashiro.
\newblock A 2.75-approximation algorithm for the unconstrained traveling tournament problem.
\newblock {\em Ann. Oper. Res.}, 218(1):237--247, 2014.

\bibitem{PM1-2}
E.~L. Lawler.
\newblock {\em Combinatorial Optimization: Networks and Matroids}.
\newblock Holt, Rinehart and Winston, New York, 1976.

\bibitem{Miyashiro}
R.~Miyashiro, T.~Matsui, and S.~Imahori.
\newblock An approximation algorithm for the traveling tournament problem.
\newblock {\em Ann. Oper. Res.}, 194:317--324, 2012.

\bibitem{ThielenWestphal2}
C.~Thielen and S.~Westphal.
\newblock Complexity of the traveling tournament problem.
\newblock {\em Theor. Comput. Sci.}, 412(4-5):345--351, 2011.

\bibitem{ThielenWestphal}
C.~Thielen and S.~Westphal.
\newblock Approximation algorithms for $\mathrm{TTP}$ (2).
\newblock {\em Math. Methods Oper. Res.}, 76:1--20, 2012.

\bibitem{XiaoKou}
M.~Xiao and S.~Kou.
\newblock An improved approximation algorithm for the traveling tournament problem with maximum trip length two.
\newblock In {\em 41st MFCS, LIPIcs}, volume~58, pages 89:1--89:14, 2016.

\bibitem{Yamaguchi}
D.~Yamaguchi, S.~Imahori, R.~Miyashiro, and T.~Matsui.
\newblock An improved approximation algorithm for the traveling tournament problem.
\newblock {\em Algorithmica}, 61(4):1077--1091, 2011.

\bibitem{ZhaoXiao}
J.~Zhao and M.~Xiao.
\newblock A further improvement on approximating $\mathrm{TTP}$-2.
\newblock In {\em 27th COCOON, LNCS}, volume 13025, pages 137--149. Springer, 2021.

\bibitem{ZhaoXiao2}
J.~Zhao and M.~Xiao.
\newblock The traveling tournament problem with maximum tour length two: A practical algorithm with an improved approximation bound.
\newblock In {\em 30th IJCAI}, pages 4206--4212, 2021.

\bibitem{ZhaoXiao3}
J.~Zhao and M.~Xiao.
\newblock Practical algorithms with guaranteed approximation ratio for $\mathrm{TTP}$ with maximum tour length two.
\newblock {\em arXiv:2212.12240}, 2022.

\bibitem{ZhaoXiaoall}
J.~Zhao and M.~Xiao.
\newblock A $5 $-approximation algorithm for the traveling tournament problem.
\newblock {\em arXiv:2309.01902}, 2023.

\bibitem{ZhaoXiaoXu}
J.~Zhao, M.~Xiao, and C.~Xu.
\newblock Improved approximation algorithms for the traveling tournament problem.
\newblock In {\em 47th MFCS, LIPIcs}, volume 241, pages 83:1--83:15, 2022.

\end{thebibliography}

\newpage


\section{Supplementary Figures}
\label{SECappA}
\begin{figure}[h]
\centering
\includegraphics[width=0.55\linewidth]{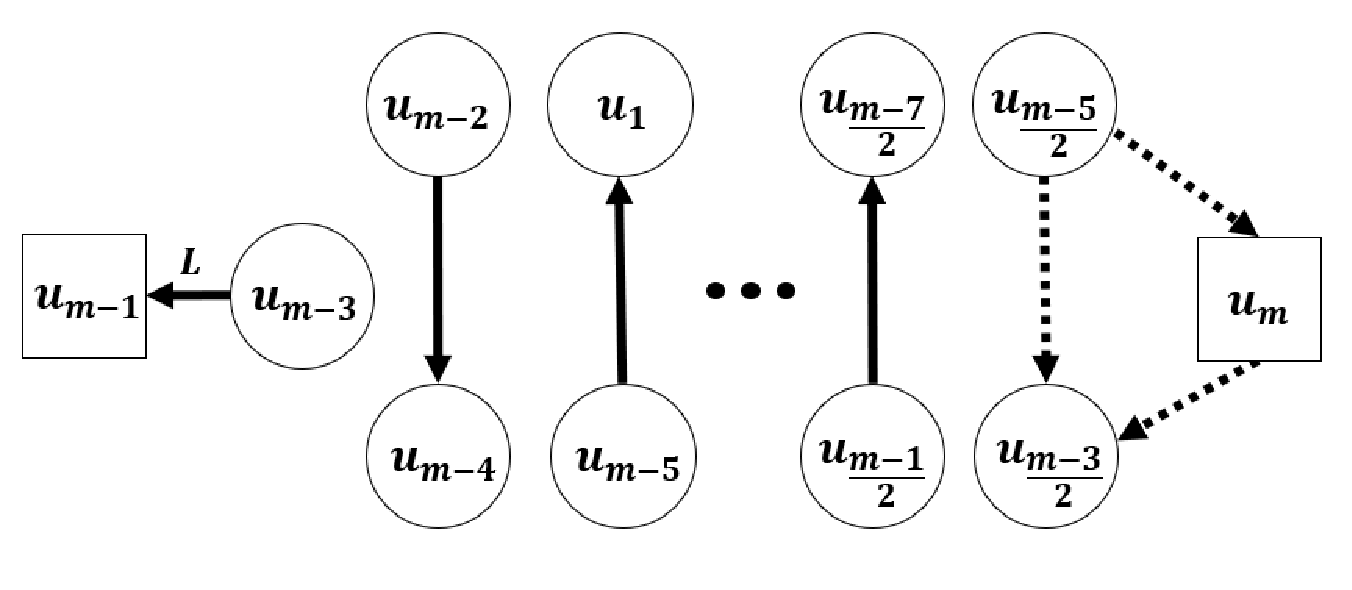}
\caption{Arrangement of the super-teams in the second slot when $n\equiv2$ (mod 8).
The edge $(u_{m-3}, u_{m-1})$ represents a left super-game.}
\label{fig3}
\end{figure}

\begin{figure}[h]
\centering
\includegraphics[width=0.55\linewidth]{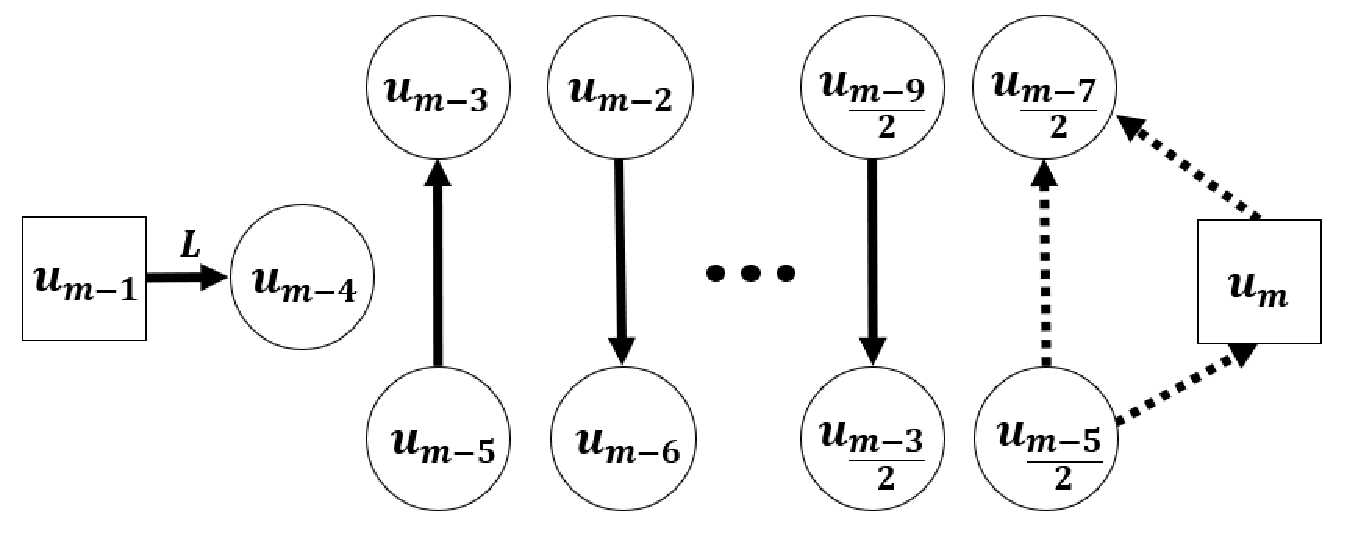}
\caption{Arrangement of the super-teams in the third slot when $n\equiv2$ (mod 8).
The edge $(u_{m-1}, u_{m-4})$ represents a left super-game.}
\label{fig4}
\end{figure}

\begin{figure}[h]
\centering
\includegraphics[width=0.55\linewidth]{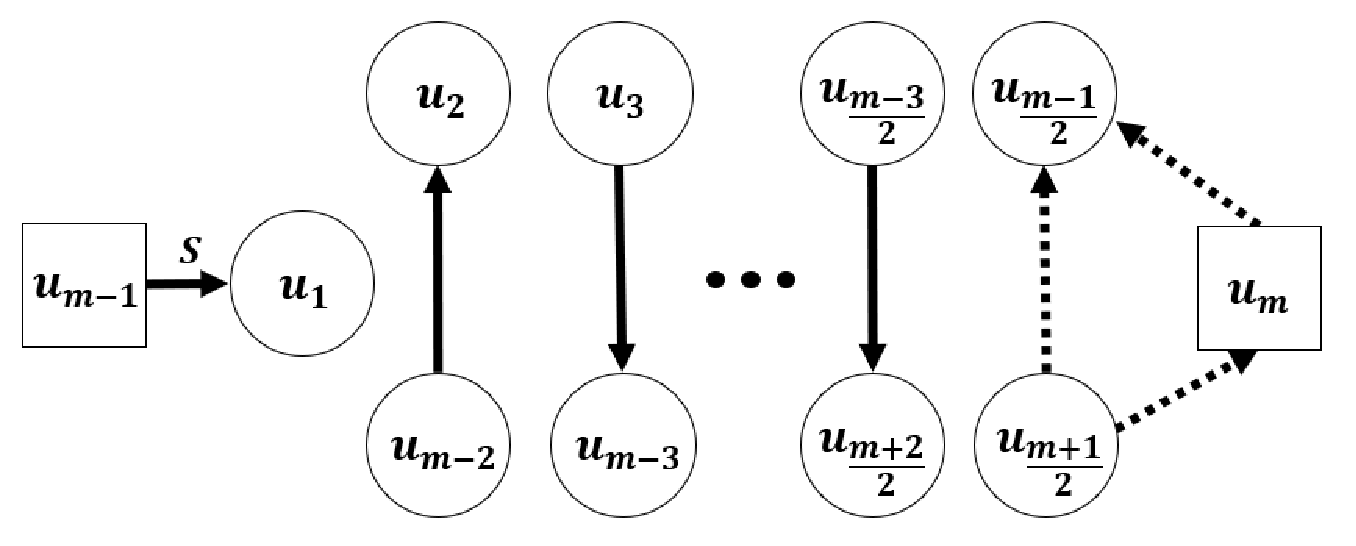}
\caption{Arrangement of the super-teams in the $(m-2)$-th slot when $n\equiv2$ (mod 8).
The edge $(u_{m-1},u_{1})$ represents a special left super-game.}
\label{fig5}
\end{figure}

\begin{figure}[h]
\centering
\includegraphics[width=0.4\linewidth]{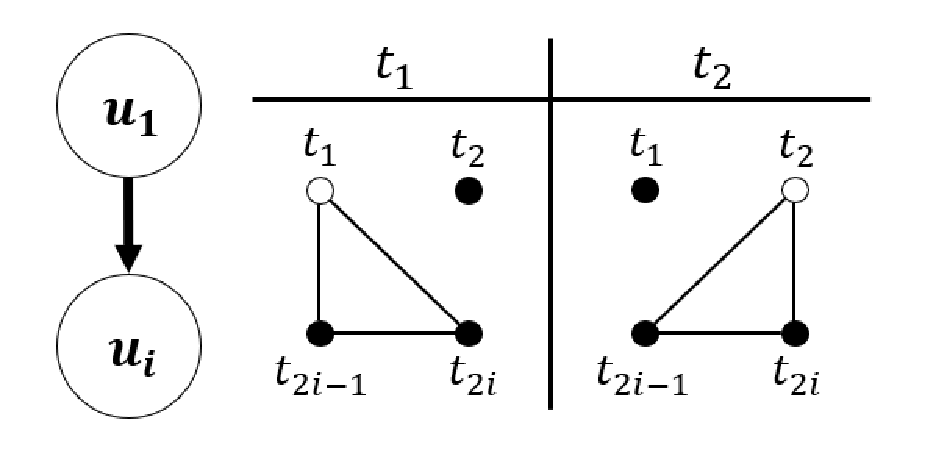}
\caption{The distance of the traveling of $t_1$ and $t_2$ in a normal super-game against super-team $u_{i}$ $(i=3,\ldots,m-3)$.
The traveling of $t_1$ and $t_2$ shown here is not extra traveling.}
\label{fig12}
\end{figure}

\begin{figure}
\centering
\includegraphics[width=0.45\linewidth]{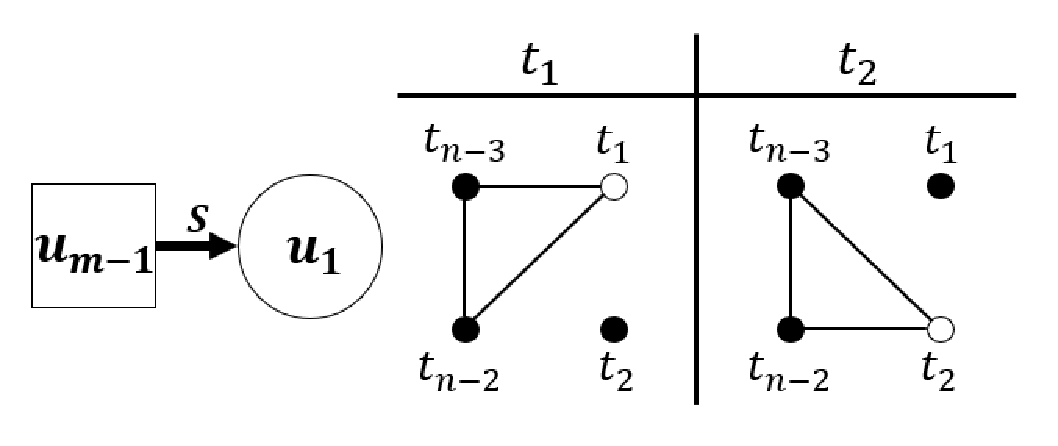}
\caption{The distance of the traveling of $t_1$ and $t_2$ in the special left super-game in the $(m-2)$-th slot.
The traveling of $t_1$ and $t_2$ shown here is not extra traveling.}
\label{fig13}
\end{figure}

\begin{figure}
\centering
\includegraphics[width=0.5\linewidth]{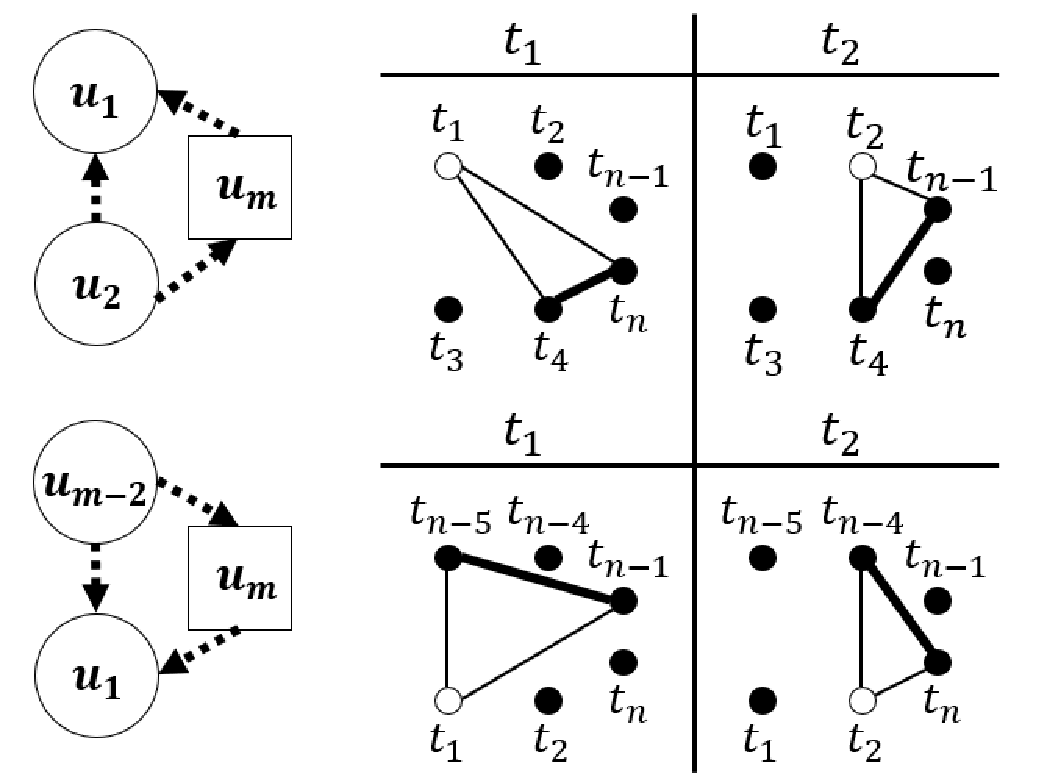}
\caption{The distance of the traveling of $t_1$ and $t_2$ in the two right super-games.
The distance of the extra traveling of $t_1$ is $d_{4,n}+d_{n-5,n-1}$, and  that of $t_2$ is $d_{4,n-1}+d_{n-4,n}$.}
\label{fig14}
\end{figure}

\begin{figure}
\centering
\includegraphics[width=0.45\linewidth]{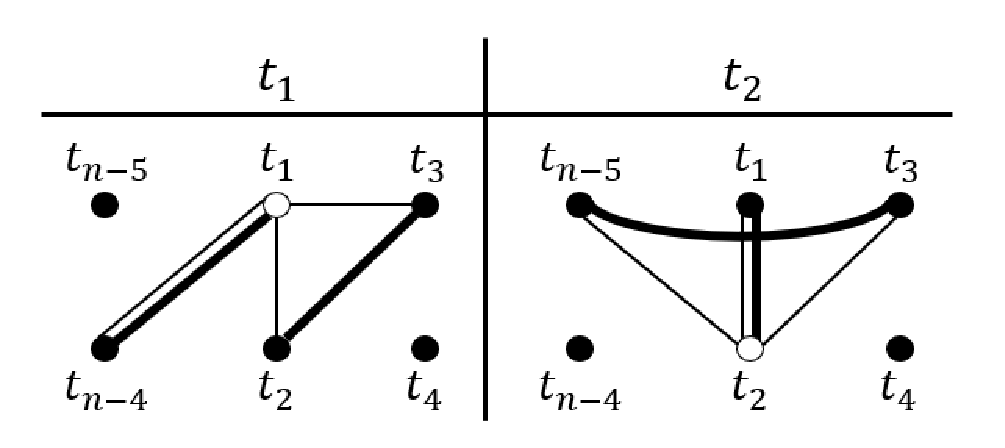}
\caption{The distance of the traveling of $t_1$ and $t_2$ in the six games in the last slot.
The distance of the extra traveling of $t_1$ is $d_{2,3}+d_{1,n-4}$, and that of $t_2$ is $d_{3,n-5}+d_{1,2}$.}
\label{fig15}
\end{figure}

\begin{figure}
\centering
\includegraphics[width=0.4\linewidth]{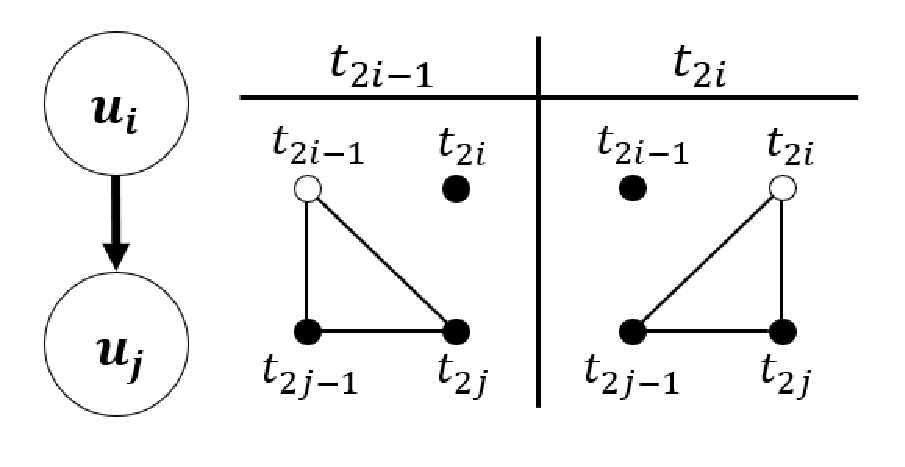}
\caption{The distance of the traveling of $t_{2i-1}$ and $t_{2i}$ $(i=2,4,\ldots,m-3)$ in a normal super-game against super-team $u_{j}$ $(j=1,\ldots,m-2,\ j\neq i-1,i,i+1)$.
The traveling of $t_{2i-1}$ and $t_{2i}$ is not extra traveling.}
\label{fig16}
\end{figure}

\begin{figure}
\centering
\includegraphics[width=0.45\linewidth]{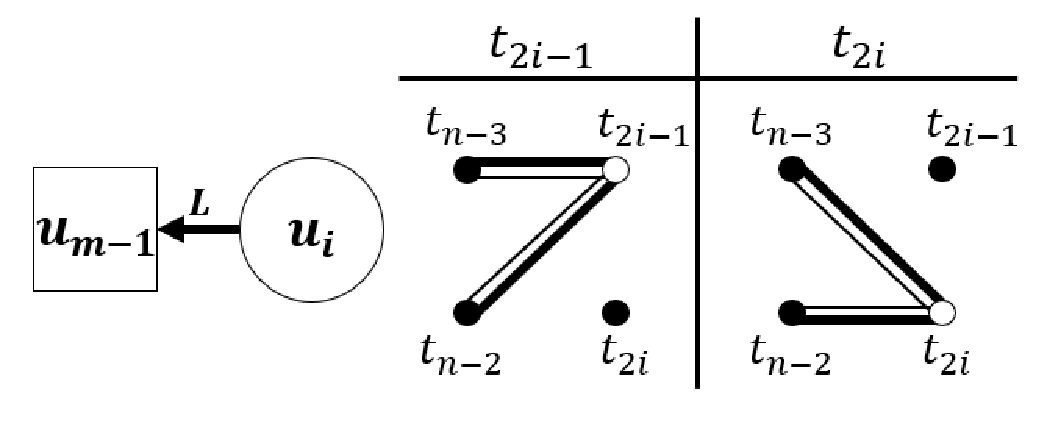}
\caption{The distance of the traveling of $t_{2i-1}$ and $t_{2i}$ $(i=2,4,\ldots,m-3)$ in the left super-game.
The distance of the extra traveling of $t_{2i-1}$ is $d_{2i-1,n-3}+d_{2i-1,n-2}$, and that of $t_{2i}$ is $d_{2i,n-3}+d_{2i,n-2}$.}
\label{fig17}
\end{figure}

\begin{figure}
\centering
\includegraphics[width=0.55\linewidth]{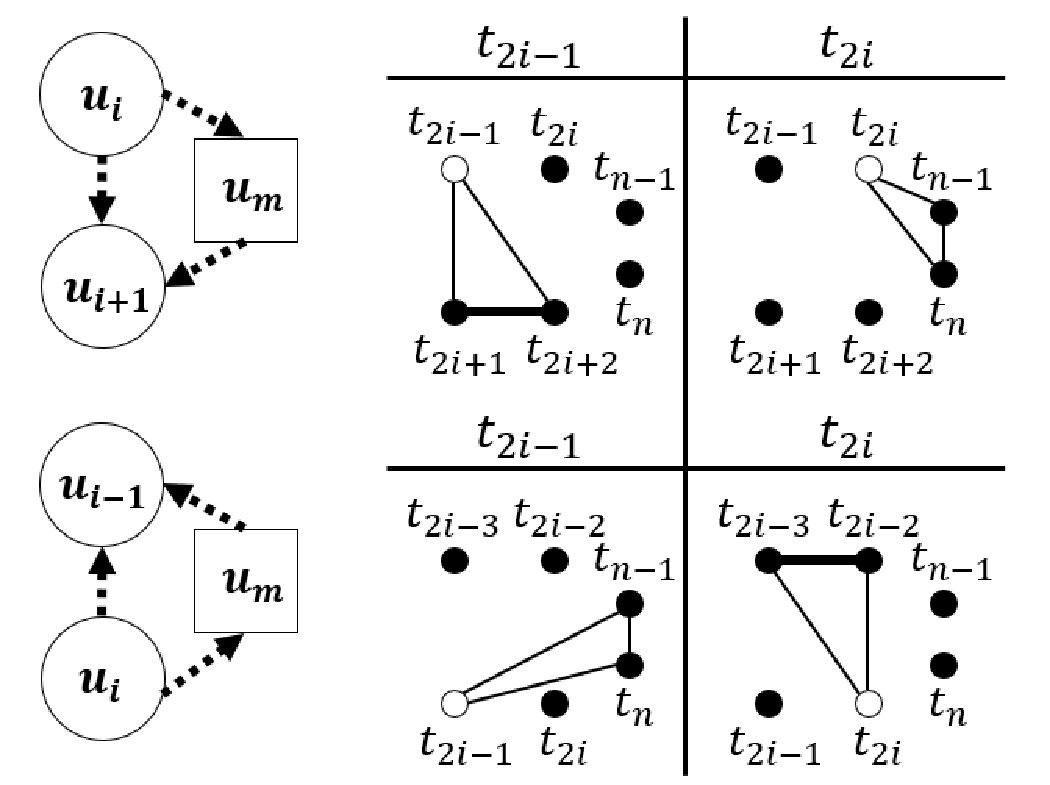}
\caption{The distance of the traveling of $t_{2i-1}$ and $t_{2i}$ $(i=2,4,\ldots,m-3)$ in the two right super-games.
The distance of the extra traveling of $t_{2i-1}$ is $d_{2i+1,2i+2}$, and that of $t_{2i}$ is $d_{2i-3,2i-2}$.}
\label{fig18}
\end{figure}

\begin{figure}
\centering
\includegraphics[width=0.45\linewidth]{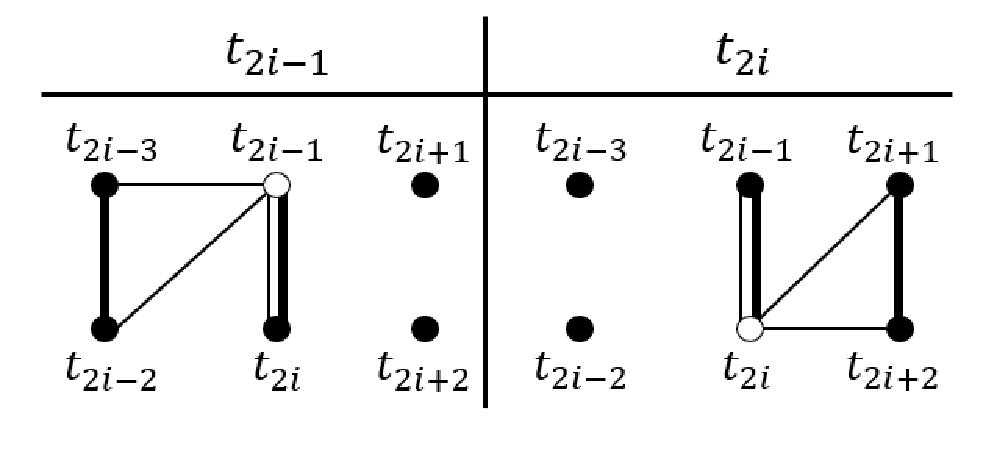}
\caption{The distance of the traveling of $t_{2i-1}$ and $t_{2i}$ $(i=2,4,\ldots,m-3)$ in the six games in the last slot.
The distance of the extra traveling of $t_{2i-1}$ is $d_{2i-1,2i}+d_{2i-3,2i-2}$, and  that of $t_{2i}$ is $d_{2i+1,2i+2}+d_{2i-1,2i}$.}
\label{fig19}
\end{figure}

\begin{figure}
\centering
\includegraphics[width=0.4\linewidth]{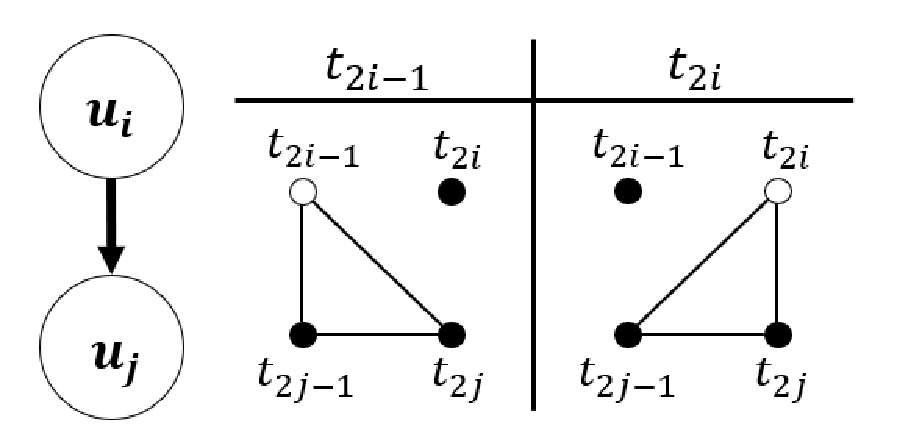}
\caption{The distance of the traveling of $t_{2i-1}$ and $t_{2i}$ $(i=3,5,\ldots,m-4)$ in a normal super-game against super-team $u_{j}$ $(j=1,\ldots,m-2,\ j\neq i-1,i,i+1)$.
The traveling of $t_{2i-1}$ and $t_{2i}$ is not extra traveling.}
\label{fig20}
\end{figure}

\begin{figure}
\centering
\includegraphics[width=0.475\linewidth]{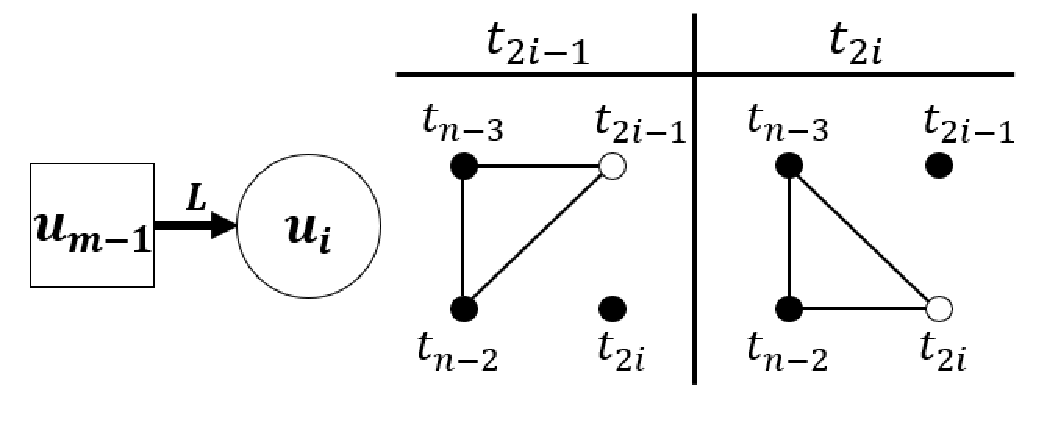}
\caption{The distance of the traveling of $t_{2i-1}$ and $t_{2i}$ $(i=3,5,\ldots,m-4)$ in the left super-game.
The traveling of $t_{2i-1}$ and $t_{2i}$ is not extra traveling.}
\label{fig21}
\end{figure}

\begin{figure}
\centering
\includegraphics[width=0.55\linewidth]{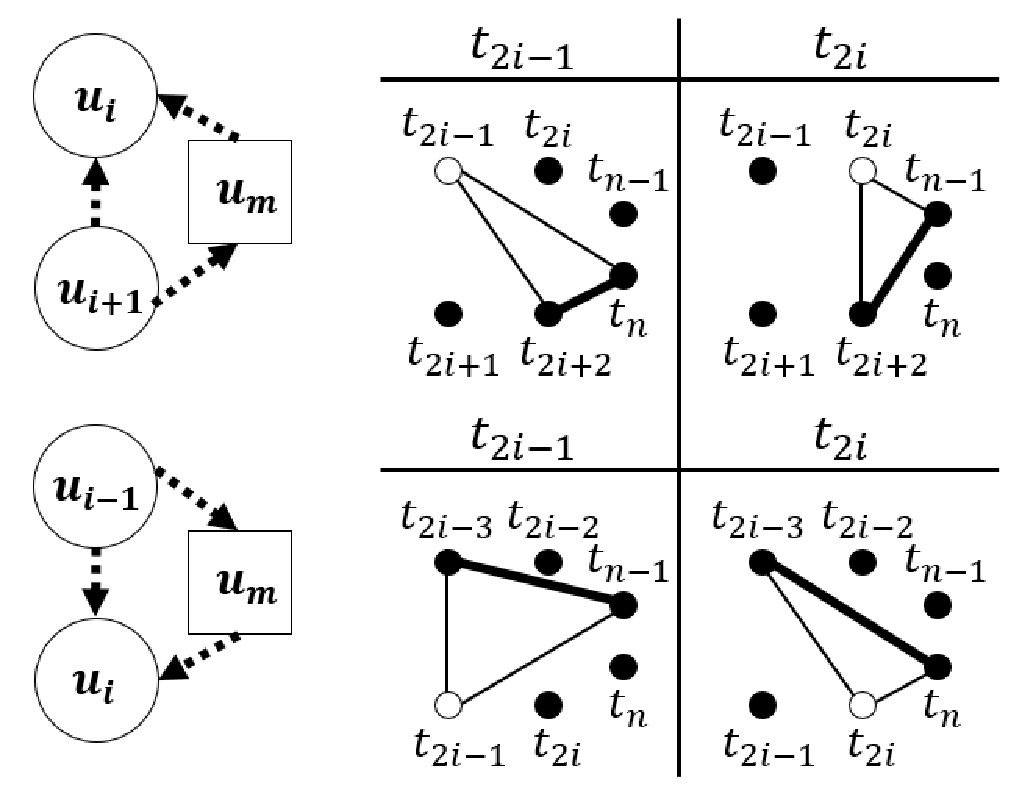}
\caption{The distance of the traveling of $t_{2i-1}$ and $t_{2i}$ $(i=3,5,\ldots,m-4)$ in the two right super-games.
The distance of the extra traveling of $t_{2i-1}$ is $d_{2i+2,n}+d_{2i-3,n-1}$, and that of $t_{2i}$ is $d_{2i+2,n-1}+d_{2i-3,n}$.}
\label{fig22}
\end{figure}

\begin{figure}
\centering
\includegraphics[width=0.5\linewidth]{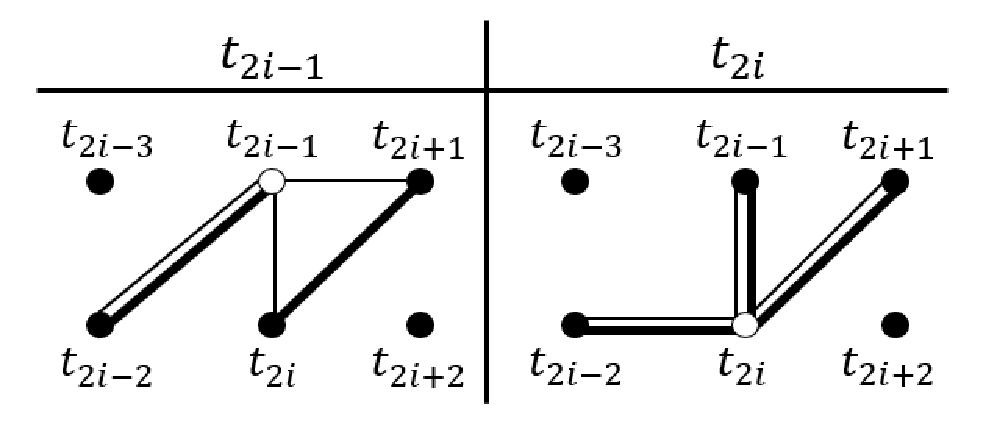}
\caption{The distance of the traveling of $t_{2i-1}$ and $t_{2i}$ $(i=3,5,\ldots,m-4)$ in the six games in the last slot.
The distance of the extra traveling of $t_{2i-1}$ is $d_{2i,2i+1}+d_{2i-2,2i-1}$, and  that of $t_{2i}$ is $d_{2i-2,2i}+d_{2i,2i+1}+d_{2i-1,2i}$.}
\label{fig23}
\end{figure}

\begin{figure}
\centering
\includegraphics[width=0.4\linewidth]{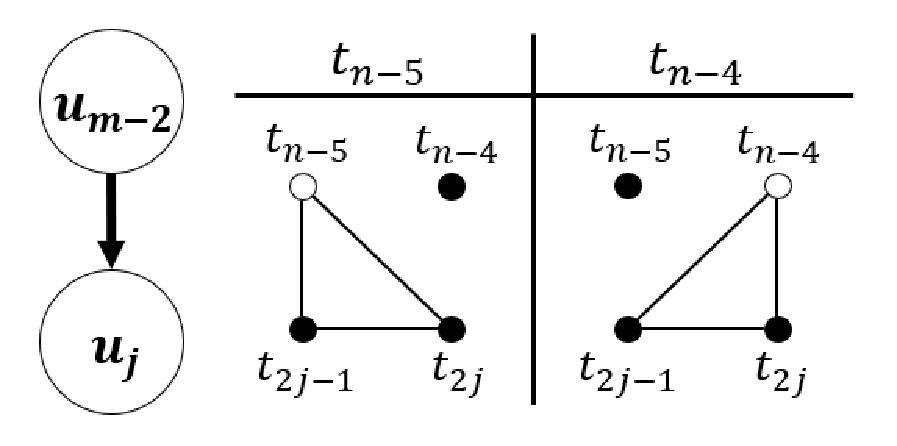}
\caption{The distance of the traveling of $t_{n-5}$ and $t_{n-4}$ in a normal super-game against super-team $u_{j}$ $(j=2,3,\ldots,m-4,m-1)$.
The traveling of $t_{n-5}$ and $t_{n-4}$ shown here is not extra traveling.}
\label{fig24}
\end{figure}

\begin{figure}
\centering
\includegraphics[width=0.5\linewidth]{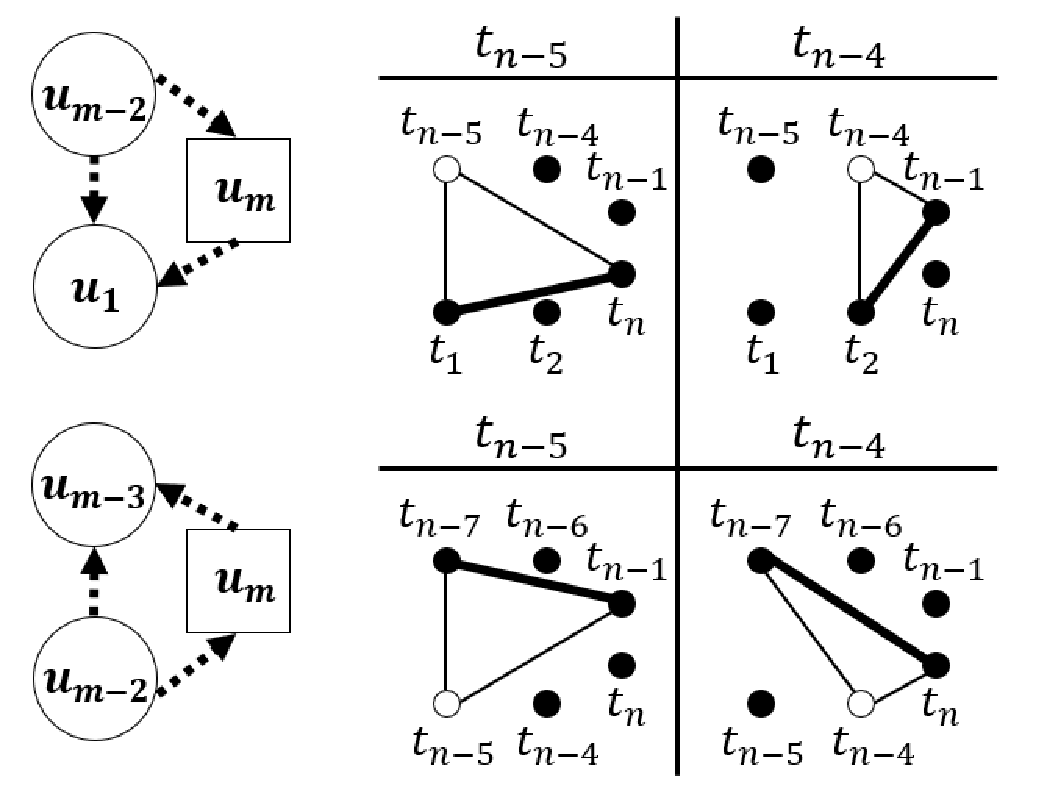}
\caption{The distance of the traveling of $t_{n-5}$ and $t_{n-4}$ in the two right super-games.
The distance of the extra traveling of $t_{n-5}$ is $d_{1,n}+d_{n-7,n-1}$, and that of $t_{n-4}$ is $d_{2,n-1}+d_{n-7,n}$.}
\label{fig25}
\end{figure}

\begin{figure}
\centering
\includegraphics[width=0.5\linewidth]{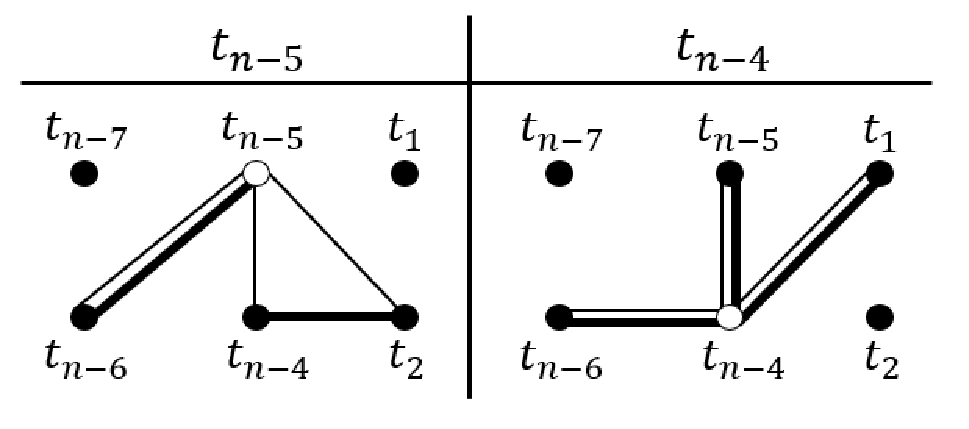}
\caption{The distance of the traveling of $t_{n-5}$ and $t_{n-4}$ in the six games in the last slot.
The distance of the extra traveling of $t_{n-5}$ is $d_{n-4,2}+d_{n-6,n-5}$, and that of $t_{n-4}$ is $d_{n-6,n-4}+d_{1,n-4}+d_{n-5,n-4}$.}
\label{fig26}
\end{figure}

\begin{figure}
\centering
\includegraphics[width=0.5\linewidth]{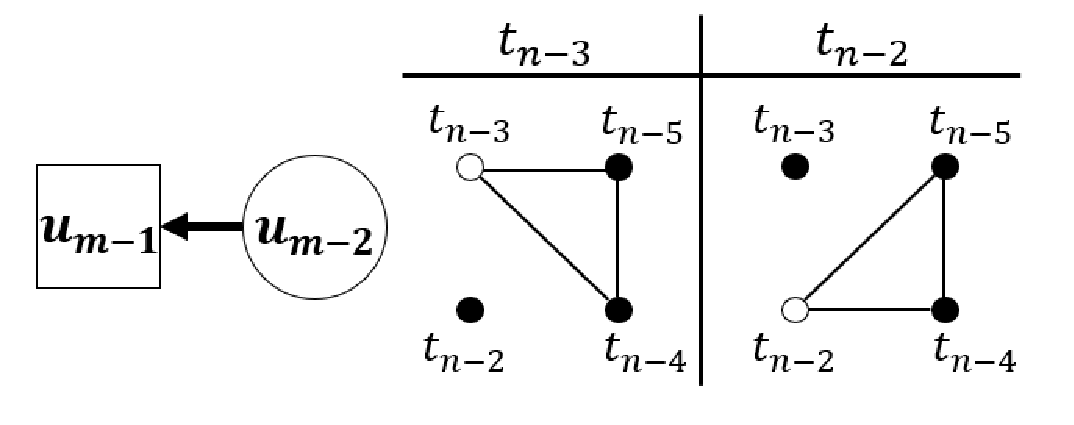}
\caption{The distance of the traveling of $t_{n-3}$ and $t_{n-2}$ in the normal super-game against the super-team $u_{m-2}$ in the first slot.
The traveling of $t_{n-3}$ and $t_{n-2}$ is not extra traveling.}
\label{fig27}
\end{figure}

\begin{figure}
\centering
\includegraphics[width=0.5\linewidth]{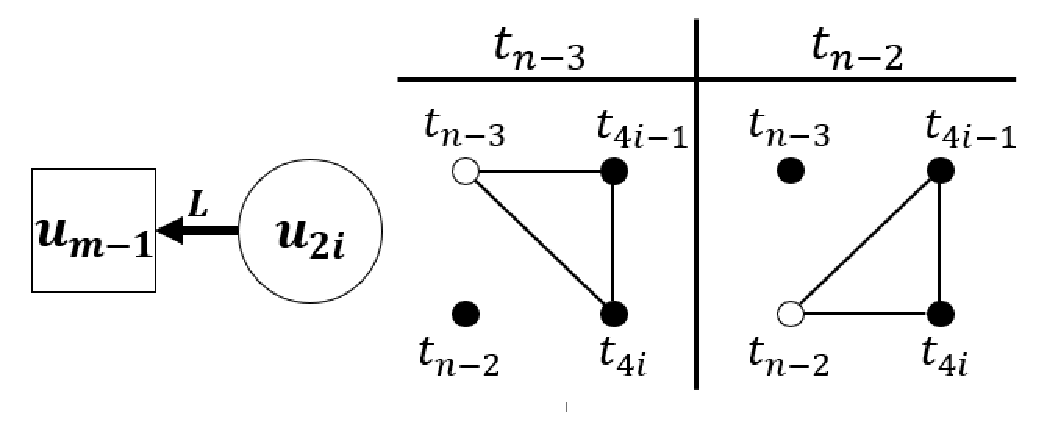}
\caption{The distance of the traveling of $t_{n-3}$ and $t_{n-2}$ in the left super-game in the $s$-th slot $(s=2,4,\ldots,m-3)$.
The traveling of $t_{n-3}$ and $t_{n-2}$ is not extra traveling.}
\label{fig28}
\end{figure}

\begin{figure}
\centering
\includegraphics[width=0.5\linewidth]{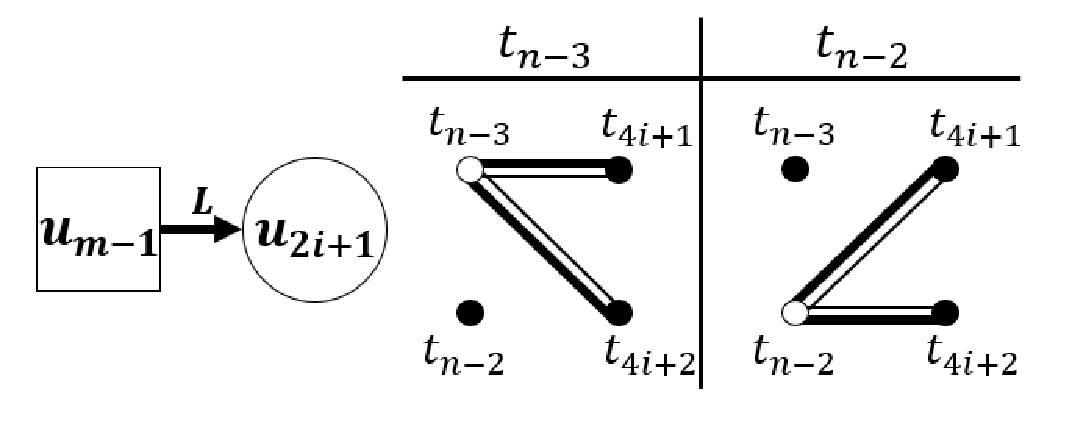}
\caption{The distance of the traveling of $t_{n-3}$ and $t_{n-2}$ in the left super-game in the $s$-th slot $(s=3,5,\ldots,m-4)$.
The distance of the extra traveling of $t_{n-3}$ is $\sum_{i=1}^{(n-10)/4}(d_{4i+1,n-3}+d_{4i+2,n-3})$, and that of $t_{n-2}$ is $\sum_{i=1}^{(n-10)/4}(d_{4i+1,n-2}+d_{4i+2,n-2})$.}
\label{fig29}
\end{figure}

\begin{figure}
\centering
\includegraphics[width=0.5\linewidth]{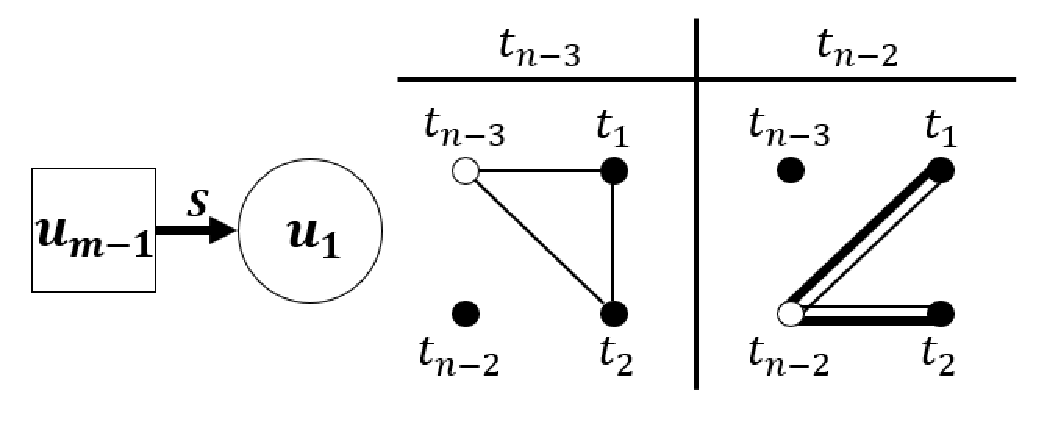}
\caption{The distance of the traveling of $t_{n-3}$ and $t_{n-2}$ in the special left super-game in the $(m-2)$-slot.
The traveling of $t_{n-3}$ is not extra traveling.
The distance of the extra traveling of $t_{n-2}$ is $d_{1,n-2}+d_{2,n-2}$.}
\label{fig43}
\end{figure}

\begin{figure}
\centering
\includegraphics[width=0.35\linewidth]{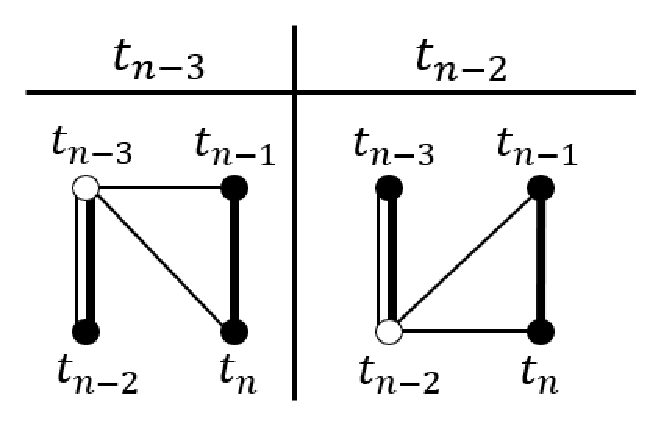}
\caption{The distance of the traveling of $t_{n-3}$ and $t_{n-2}$ in the six games in the last slot.
The distance of the extra traveling of $t_{n-3}$ is $d_{n-2,n-3}+d_{n,n-1}$, and that of $t_{n-2}$ is $d_{n-1,n}+d_{n-3,n-2}$.}
\label{fig30}
\end{figure}

\begin{figure}
\centering
\includegraphics[width=0.55\linewidth]{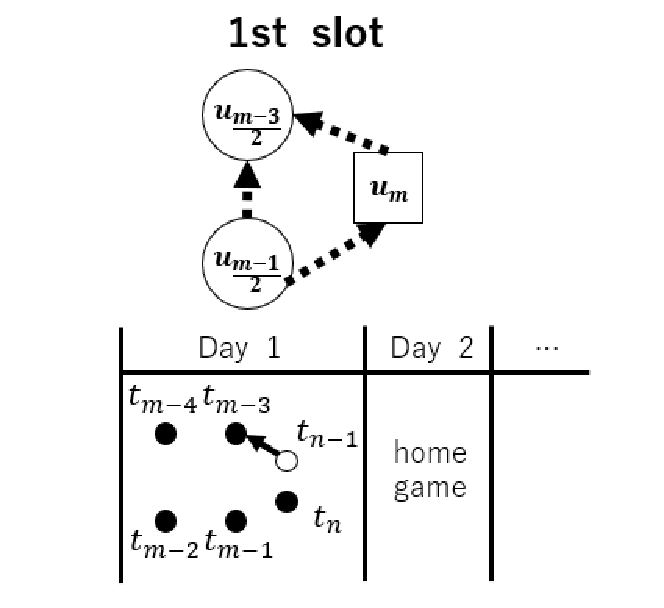}
\caption{The games of $t_{n-1}$ in the right super-game on Days 1 and 2 in the first slot.
Team $t_{n-1}$ travels from its home venue to that of $t_{m-3}$, and then travels back to its home venue.
Hence the distance of the extra traveling of $t_{n-1}$ is $d_{m-3,n-1}$.}
\label{fig41}
\end{figure}

\begin{figure}
\centering
\includegraphics[width=0.65\linewidth]{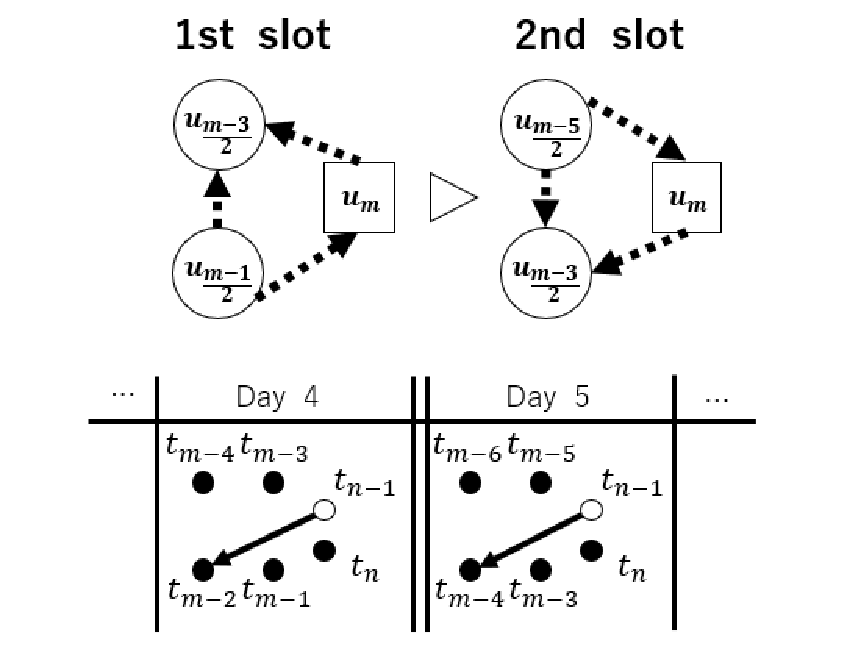}
\caption{The games of $t_{n-1}$ on Day 4 in the first slot and on Day 5 in the second slot.
Team $t_{n-1}$ travels from the home venue of $t_{m-2}$ to that of $t_{m-4}$ between Days 4 and 5.
Hence the distance of the extra traveling of $t_{n-1}$ is $d_{m-4,m-2}$.}
\label{fig44}
\end{figure}

\begin{figure}
\centering
\includegraphics[width=1.0\linewidth]{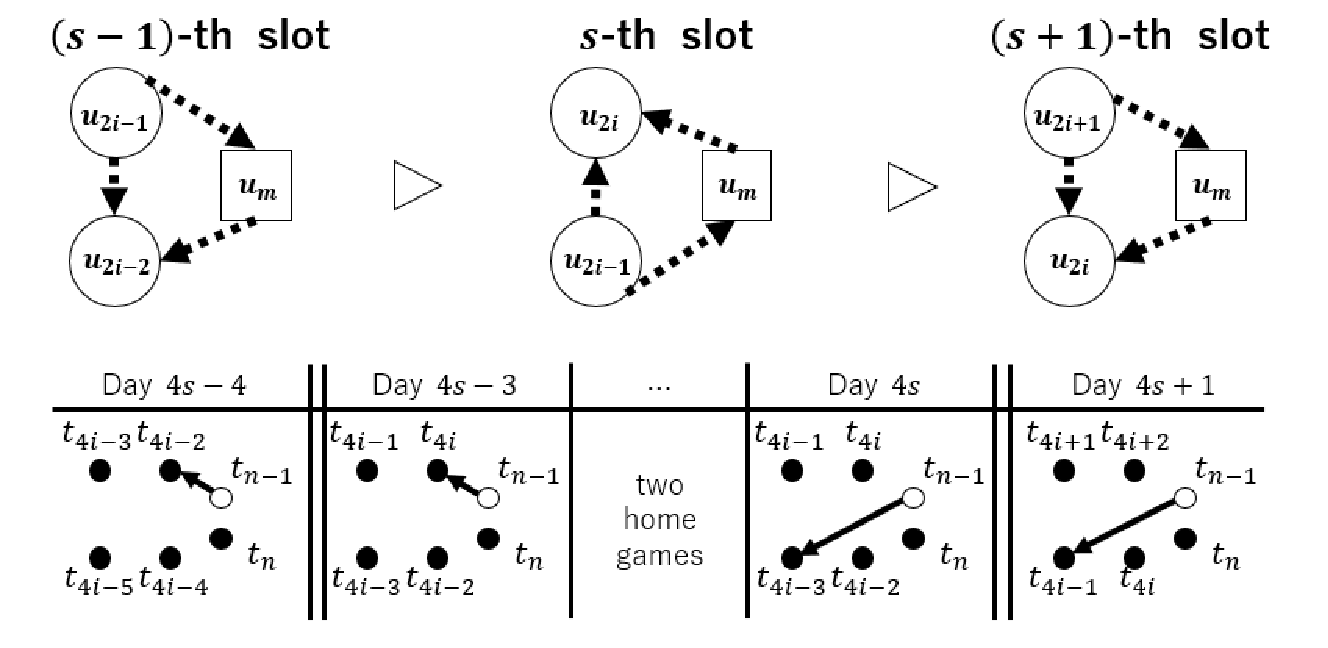}
\caption{The games of $t_{n-1}$ in the three right super-games in $(s-1)$-th, $s$-th, and $(s+1)$-th slots $(s=3,5,\ldots,(m-3)/2)$.
Team $t_{n-1}$ travels from the home venue of $t_{4i-2}$ to the home venue of $t_{4i}$ between Days $4s-4$ and $4s-3$, and from the home venue of $t_{4i-3}$ to that of $t_{4i-1}$ between Days $4s$ and $4s+1$.
Hence the distance of the extra traveling of $t_{n-1}$ is $d_{4i-3,4i-1}+d_{4i-2,4i}$ $(i=1,\ldots,(m-5)/4)$, and the total distance of these extra travelings is $\sum_{i=1}^{(m-5)/4}(d_{4i-3,4i-1}+d_{4i-2,4i})$.}
\label{fig31-a}
\end{figure}

\begin{figure}
\centering
\includegraphics[width=1.0\linewidth]{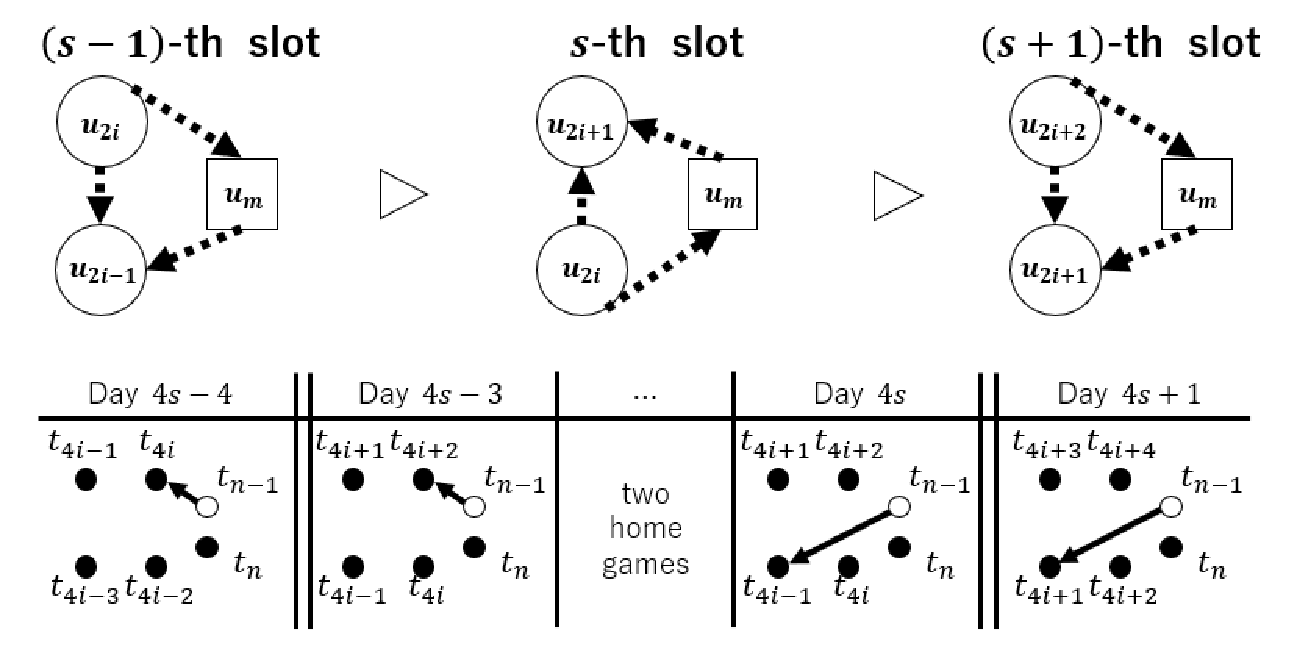}
\caption{The games of $t_{n-1}$ in the three right super-games in $(s-1)$-th, $s$-th, and $(s+1)$-th slots $(s=(m+1)/2,(m+3)/2,\ldots,m-4)$.
Team $t_{n-1}$ travels from the home venue of $t_{4i}$ to that of $t_{4i+2}$ between Days $4s-4$ and $4s-3$, and from the home venue of $t_{4i-1}$ to that of $t_{4i+1}$ between Days $4s$ and $4s+1$.
Hence the distance of the extra traveling of $t_{n-1}$ is $d_{4i-1,4i+1}+d_{4i,4i+2}$ $(i=(m+3)/4,\ldots,(m-3)/2)$, and the total distance of these extra travelings is $\sum_{i=(m+3)/4}^{(m-3)/2}(d_{4i-1,4i+1}+d_{4i,4i+2})$.}
\label{fig31-b}
\end{figure}

\begin{figure}
\centering
\includegraphics[width=0.65\linewidth]{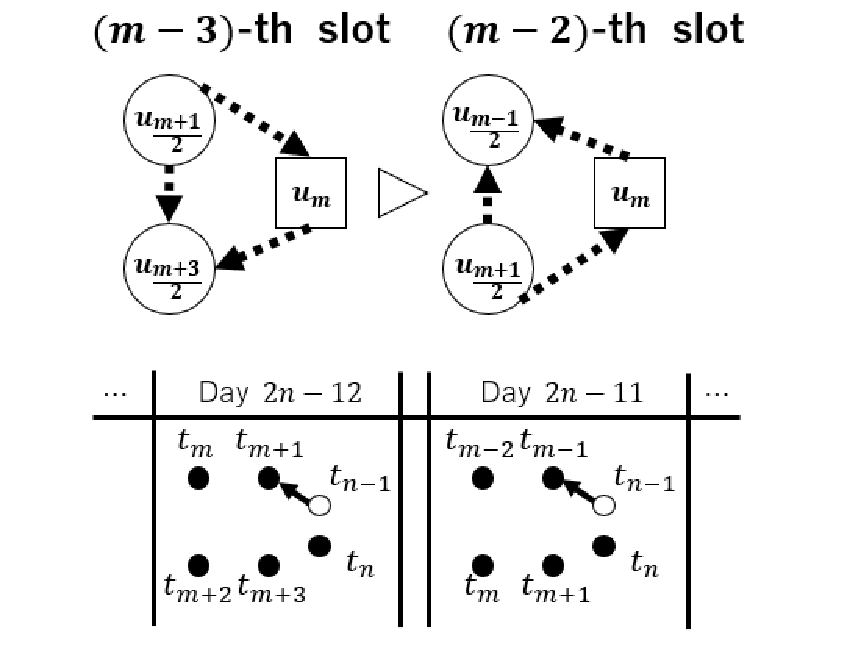}
\caption{The games of $t_{n-1}$ on Day $2n-12$ in the $(m-3)$-th slot and on Day $2n-11$ in the $(m-2)$-th slot.
Team $t_{n-1}$ travels from the home venue of $t_{m+1}$ to that of $t_{m-1}$ between Day $2n-12$ and Day $2n-11$.
Hence the distance of the extra traveling of $t_{n-1}$ is $d_{m-1,m+1}$.}
\label{fig45}
\end{figure}

\begin{figure}
\centering
\includegraphics[width=0.7\linewidth]{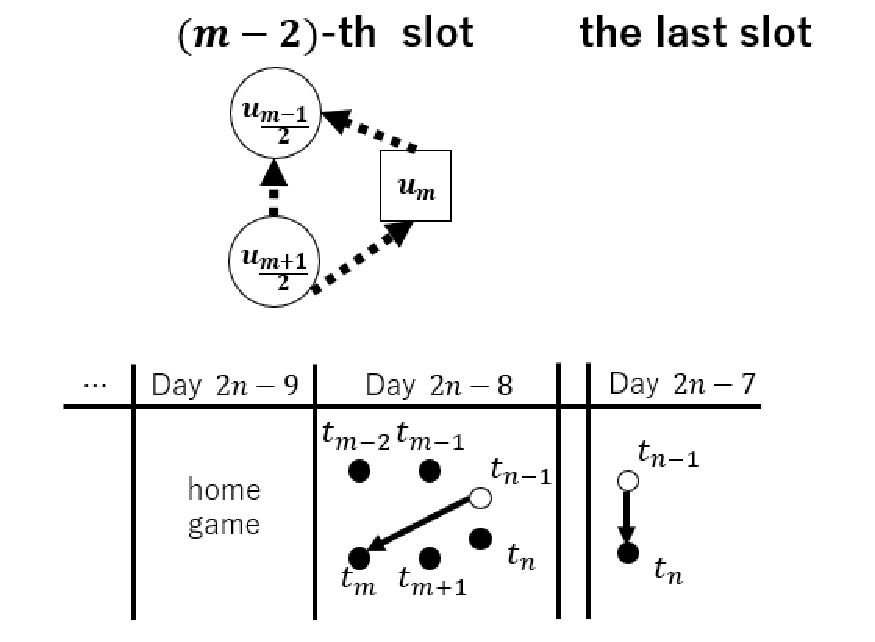}
\caption{The games of $t_{n-1}$ on Day $2n-8$ in the $(m-2)$-th slot and on Day $2n-7$ in the last slot.
Team $t_{n-1}$ travels from the home venue of $t_{m}$ to that of $t_{n}$ between Days $2n-8$ and $2n-7$.
Hence the distance of the extra traveling of $t_{n-1}$ is $d_{m,n}$.}
\label{fig42}
\end{figure}

\begin{figure}
\centering
\includegraphics[width=0.65\linewidth]{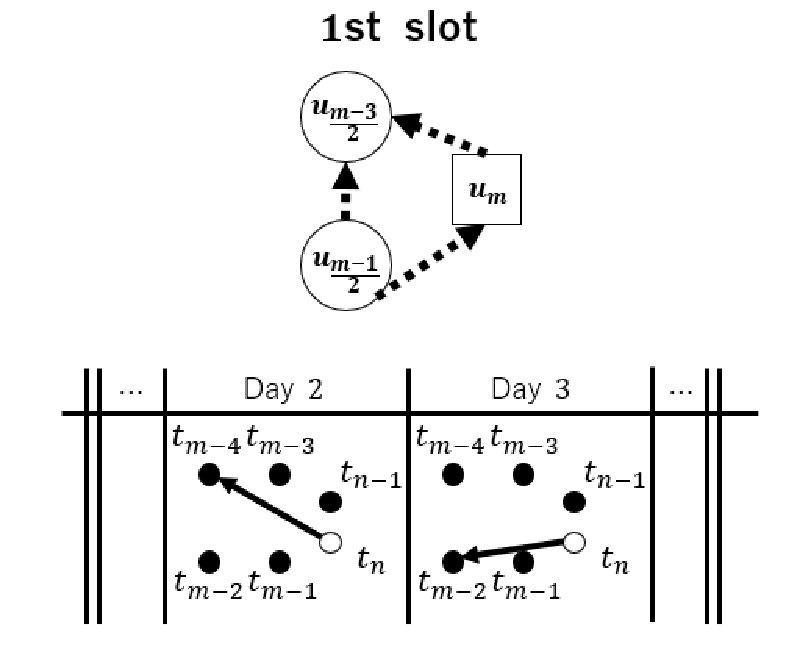}
\caption{The games of $t_{n}$ on Days 2 and 3 in the first slot.
Team $t_{n}$ travels from the home venue of $t_{m-4}$ to that of $t_{m-2}$ between Days 2 and 3.
Hence the distance of the extra traveling of $t_{n}$ is $d_{m-4,m-2}$.}
\label{fig46}
\end{figure}

\begin{figure}
\centering
\includegraphics[width=0.95\linewidth]{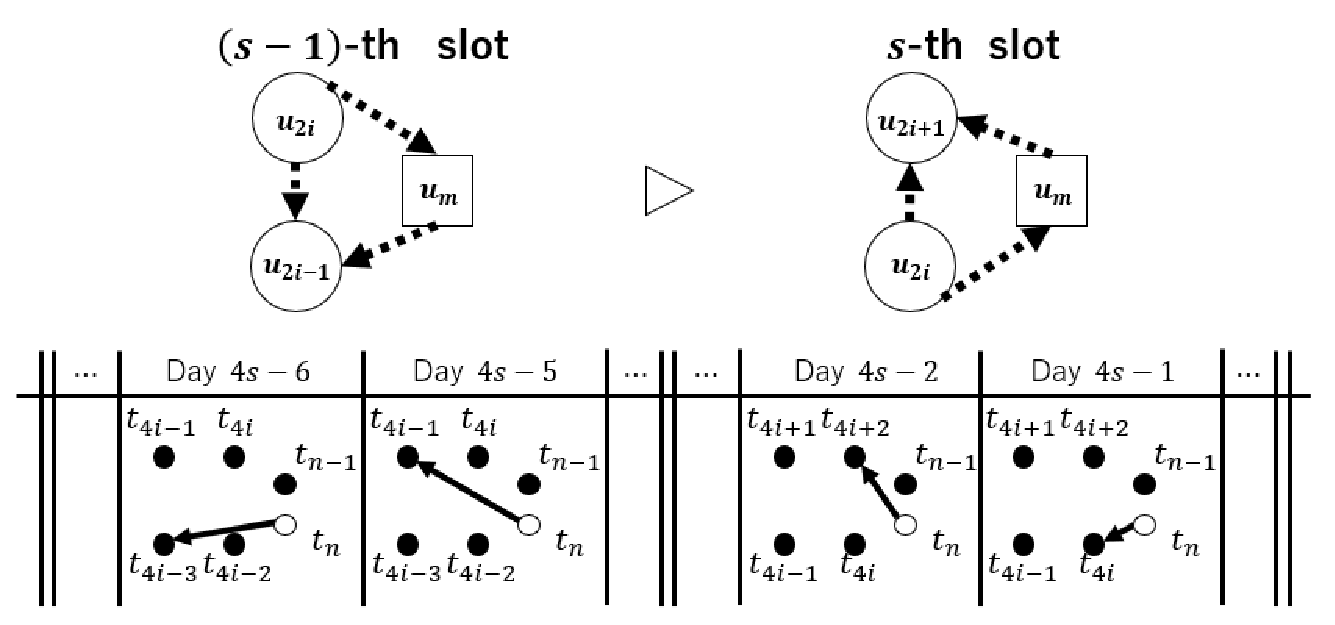}
\caption{The games of $t_{n}$ in the two right super-games in $(s-1)$-th and $s$-th slots $(s=3,5,\ldots,(m-3)/2)$.
Team $t_{n}$ travels from the home venue of $t_{4i-3}$ to that of $t_{4i-1}$ between Days $4s-6$ and $4s-5$, and from the home venue of $t_{4i+2}$ to that of $t_{4i}$ between Days $4s-2$ and $4s-1$.
Hence the distance of the extra traveling of $t_{n-1}$ is $d_{4i-3,4i-1}+d_{4i,4i+2}$ $(i=1,\ldots,(m-5)/4)$, and the total distance of these extra travelings is $\sum_{i=1}^{(m-5)/4}(d_{4i-3,4i-1}+d_{4i,4i+2})$.}
\label{fig47}
\end{figure}

\begin{figure}
\centering
\includegraphics[width=0.6\linewidth]{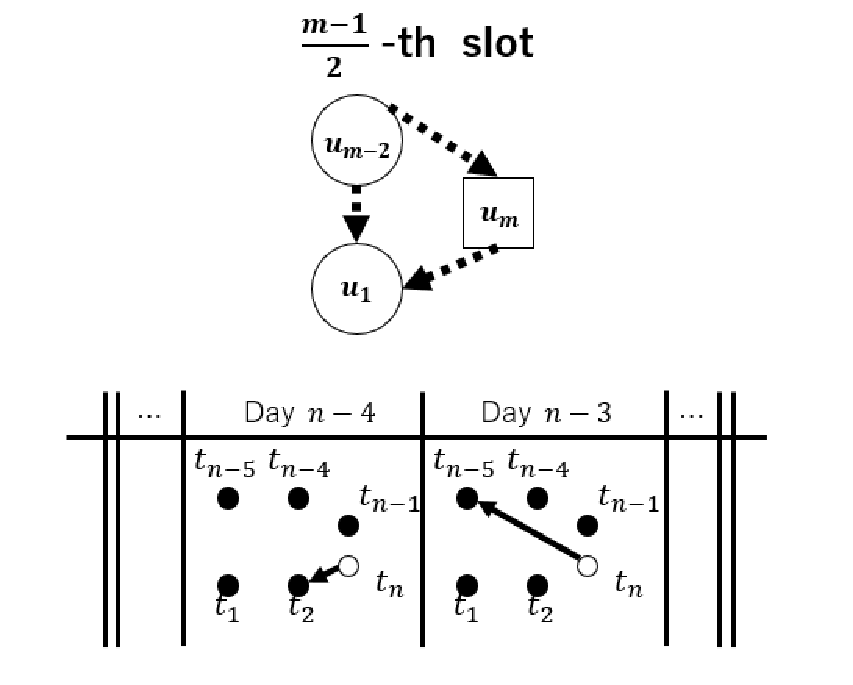}
\caption{The games of $t_{n}$ on Days $n-4$ and $n-3$ in the $(m-1)/2$-th slot.
Team $t_{n}$ travels from the home venue of $t_{2}$ to that of $t_{n-5}$ between Days $n-4$ and $n-3$.
Hence the distance of the extra traveling of $t_{n}$ is $d_{2,n-5}$.}
\label{fig35}
\end{figure}

\begin{figure}
\centering
\includegraphics[width=0.9\linewidth]{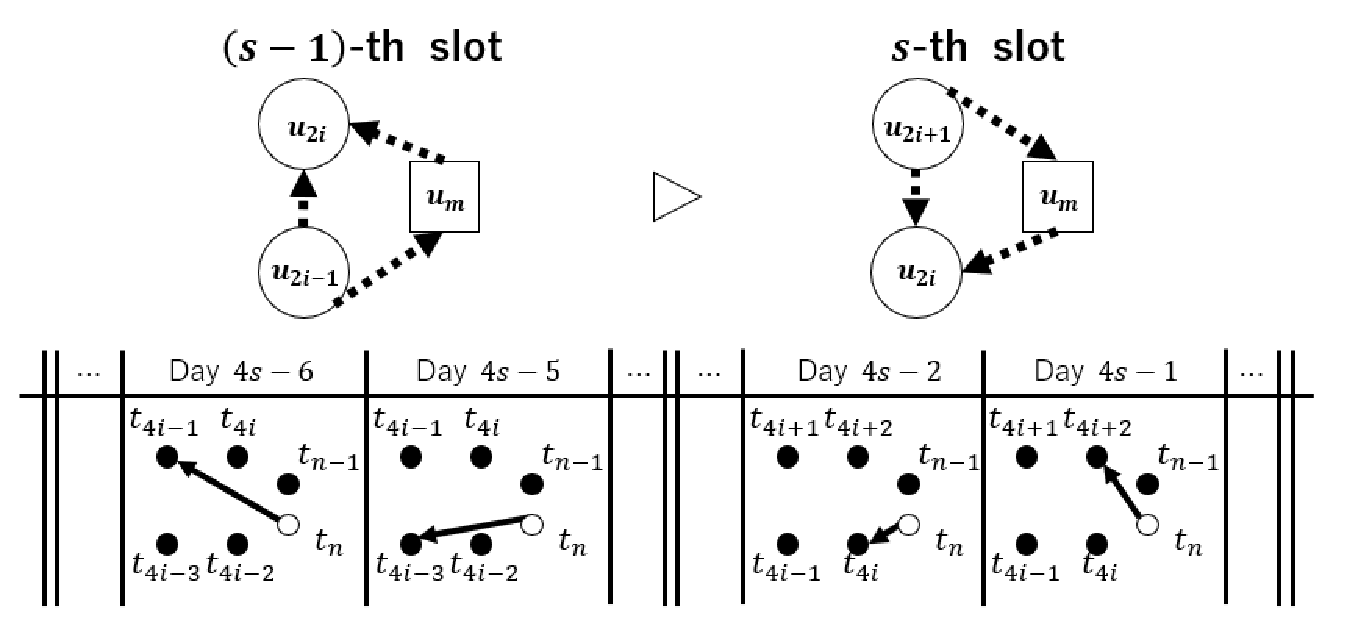}
\caption{The games of $t_{n}$ in the two right super-games in $(s-1)$-th and $s$-th slots $(s=(m+3)/2,(m+5)/2,\ldots,m-3)$.
Team $t_{n}$ travels from the home venue of $t_{4i-1}$ to that of $t_{4i-3}$ between Days $4s-6$ and $4s-5$, and from the home venue of $t_{4i}$ to that of $t_{4i+2}$ between Days $4s-2$ and $4s-1$.
Hence the distance of the extra traveling of $t_{n-1}$ is $d_{4i-3,4i-1}+d_{4i,4i+2}$ $(i=(m+3)/4,\ldots,(m-3)/2)$, and the total distance of these extra travelings is $\sum_{i=(m+3)/4}^{(m-3)/2}(d_{4i-3,4i-1}+d_{4i,4i+2})$.}
\label{fig48}
\end{figure}

\begin{figure}
\centering
\includegraphics[width=0.6\linewidth]{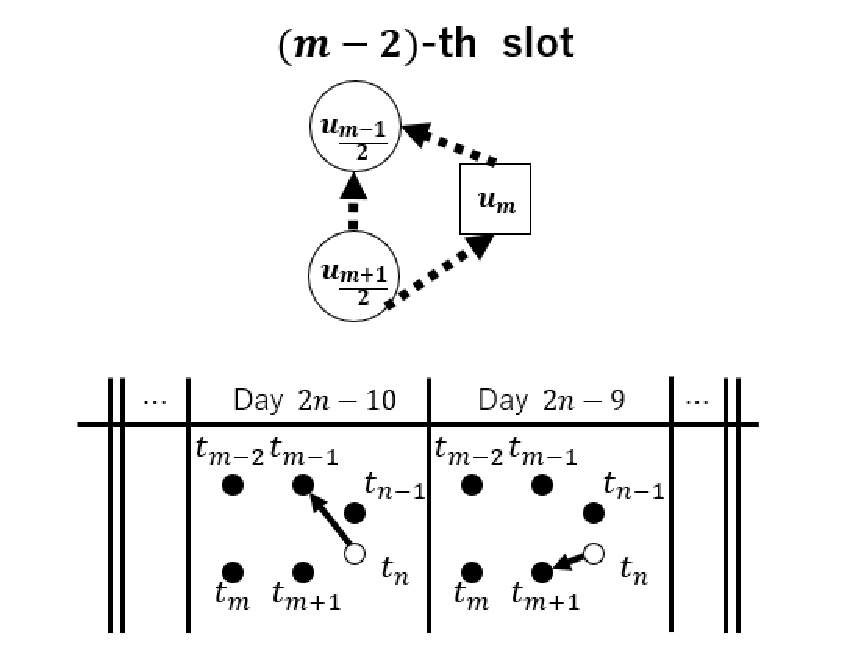}
\caption{The games of $t_{n}$ on Days $2n-10$ and $2n-9$ in the $(m-2)$-th slot.
Team $t_{n}$ travels from the home venue of $t_{m-1}$ to that of $t_{m+1}$ between Days $2n-10$ and $2n-9$.
Hence the distance of the extra traveling of $t_{n}$ is $d_{m-1,m+1}$.}
\label{fig49}
\end{figure}

\begin{figure}
\centering
\includegraphics[width=0.35\linewidth]{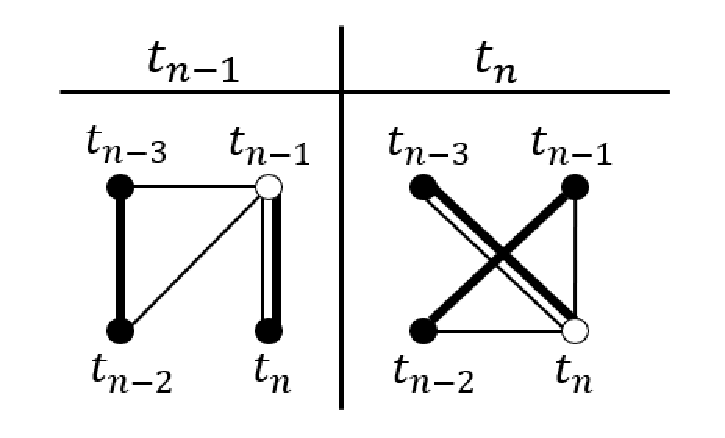}
\caption{The distance of the traveling of $t_{n-1}$ and $t_{n}$ in the six games in the last slot.
The distance of the extra traveling of $t_{n-1}$ is $d_{n-3,n-2}+d_{n-1,n}$, and that of $t_n$ is $d_{n-2,n-1}+d_{n-3,n}$.}
\label{fig36}
\end{figure}

We show Figures \ref{fig3}--\ref{fig5} and Figures \ref{fig12}--\ref{fig36}, which are omitted in Section \ref{Construction} and Section \ref{approximation ratio},  respectively.

\end{document}